# A novel analysis method for calculating nonlinear Frequency Response Functions

D. Di Maio[1]

[1] University of Twente, Faculty of Engineering Technology, MS[3] department
De Horst 2, 7522 LW, Enschede, the Netherlands
e-mail: **d.dimaio@utwente.nl**

## Abstract

The Frequency Response Functions (*FRFs*) are the most widely used functions to characterise the dynamic behaviour of structures. The natural frequencies and damping behaviour can be easily and quickly detected from a Bode diagram. The modal properties of *FRFs* can be evaluated using modal analysis methods, and as the last step, frequency response models can synthesise response functions to verify the robustness of the modal parameters identified by the analysis. The circularity between 1) measurement, 2) identification, 3) regeneration and 4) comparison is ensured on the assumption that transfer functions are measured under linear vibrations, even though mechanical systems are intrinsically non-linear. Some sources of nonlinearity might be excited and revealed, and others not for various reasons. Anyhow, it is unavoidable to measure non-linear vibrations when vibration tests are executed at various levels of excitation forces. Eventually, linear and non-linear vibrations are processed to obtain linear and non-linear *FRF*s. The linear *FRF*s are processed using the existing identification methods. The non-linear *FRF*s are archived or blandly processed to evaluate the level and the type of nonlinearity, such as hardening or softening behaviour.

This research aims to (i) formulate a new analysis method to generate nonlinear frequency responses and (ii) formulate a new identification method for extracting amplitude-dependent modal parameters. The first objective will demonstrate that a *nonlinear frequency response* surface generated by linear *FRFs* is the solution space of nonlinear *FRFs*. The second objective will demonstrate that a linear modal analysis method called line-fit, based on the Dobson formulation, allows extracting amplitude-dependent modal parameters from non-linear *FRFs*. Combining the results of the two research objectives, one can regenerate and compare measured and synthesised non-linear frequency response functions.

## 1 Introduction

Modal testing and analysis are well-established practices for measuring and analysing frequency response functions. The modal parameters identified from the analysis can be used to generate a simple mathematical response model or validate and update a structural model generated by finite element methods. There are four essential steps in the modal analysis practice. The first one is the measurement of frequency response frequencies, *FRFs*, which are the ratio between response and stimulus, measured under steady-state, random or transient vibrations. The vibrations are acquired and processed by linear operators, such as the Fast Fourier Transform (*FFT*), to calculate frequency response spectra. The second step is the modal analysis, which identifies the modal parameters from the *FRFs*. The process can be either via time- or frequency-domain methods. The third step is to generate or synthesise *FRFs* using linear frequency response models with the modal parameters identified in the second step. The final fourth step is the comparison of the measured and synthesised *FRFs*, a process that ensures the reliability of the modal parameters. Nowadays, modal analysis toolsets are implemented in several commercial and open-source software suites and used in several engineering applications. The circularity between 1) measurement, 2) identification, 3) regeneration and 4) comparison ensures that the analysis process delivers robust and reliable modal parameters. Ewins [1], Maia [2], Avitable [3], to name a few, worked extensively on modal analysis, and they cast this modus operandi on solid foundations for the future generation of practitioners to use it reliably.

Any mechanical system is intrinsically nonlinear by nature, and therefore, the circularity described earlier breaks as soon as nonlinear vibrations are excited and measured. Under such nonlinear conditions, the *FRFs* will become more and more distorted. Such distortions can become visible around the resonances, which tend to skew toward higher or lower frequencies. There are also situations where the resonance distortion is not visible. Therefore, any modal test practitioner must often conduct linearity checks before running any test. Worden wrote a textbook describing the nonlinear structural dynamics [4], in which several types of nonlinearities are modelled in the frequency and time domain, and experimental techniques for characterising nonlinearities are also presented. Although nonlinear dynamics is a science that covers many applications, from engineering to physics and mathematics, this manuscript narrows its focus to the steady state nonlinear vibrations, which are only analysed using *FRFs*.

The frequency response functions are widely used in engineering to describe the dynamics of systems. Test campaigns are expensive both in terms of money and time, and once these tests are planned and committed, the practitioners will measure the data using industry standards. The linear *FRFs* are processed, and the results will stream into the project workflow. The nonlinear ones might be blandly used for assessing the nonlinearity of a structure and then archived because commercial modal analysis software cannot handle them properly. The engineering challenge is to overcome this limitation, and this manuscript postulates that:

*A nonlinear frequency response surface, generated by a waterfall of linear FRFs, is the solution space of nonlinear FRFs, which can be evaluated by any force plane cutting across the response surface.*

The hypothesis must address the following scientific objectives:

a. The primary scientific objective is to develop a new mathematical formulation to (i) analyse a *nonlinear frequency response* surface and (ii) calculate nonlinear frequency response functions.
b. The secondary scientific objective is to develop a new modal analysis tool to calculate amplitude-dependent modal parameters using one *FRF* rather than several ones.

This manuscript will present three sections of research work. The first section will focus on developing a mathematical formulation for evaluating nonlinear transfer functions for one and two degrees of freedom systems. It will demonstrate that a *nonlinear frequency response* surface generated by a waterfall of linear *FRFs* can be analysed to calculate nonlinear *FRFs*. Such a surface can be yielded either by accessing i) nonlinear force-displacement relationships or ii) modal parameters amplitude-dependent. Finally, the nonlinear *FRF*s evaluated by the proposed method are compared to nonlinear *FRFs* calculated by numerical integration and the Harmonic Balance Method. The Appendix presents two simple case studies of structures modelled by the finite element method, showing how to use the proposed analysis to calculate nonlinear *FRFs*. The second section will present a novel modal analysis method based on the Dobson formulation [5], which allows calculating amplitude-dependent modal parameters from one nonlinear *FRF*. The third section will present three case studies to validate the new formulation proposed: one experimental case based on a lap joint and two experimental cases based on a composite blade. The formulation proposed in the paper's first section will allow the calculation of nonlinear *FRFs* using the amplitude-dependent modal parameters measured by the experiments.

It is important to stress that the proposed formulation and application of the method are currently developed on the following limitations.

1) Experimental and synthesised measurements are yielded using stepped sine tests under steady state conditions.
2) The *FRFs*, calculated from the measurements, are the ratio between the response and stimulus at the fundamental drive frequency, and therefore, higher-order harmonics are neglected.
3) The resonances are considered well-isolated, meaning that modal contribution from neighbouring modes can be neglected.
4) The manuscript focuses on smooth nonlinearities.

## 2    Response function model for nonlinear steady-state vibrations

The primary scientific objective of this section is to demonstrate the hypothesis postulated in the introduction. The investigation starts by revisiting the relationship between force and displacement, which contains an element to understand why linear *FRFs* can generate a *nonlinear frequency response* surface.

This section will only treat a cubic nonlinear stiffness, leaving the nonlinear damping out of the scope of this demonstration. The nonlinear *FRF* yielded by the proposed new analysis method (NM) will be compared to those calculated using the Harmonic Balance Method and numerical integration method. The section will focus on the ONE- and TWO-Degrees of Freedom systems.

## 2.1 Force-displacement relationship for linear system

An idealised spring, subjected to a static force, will extend or compress about a displacement ($X$). It is custom to write the force-displacement relationship as in equation (1), where $k$, is the stiffness coefficient. An alternative form can be used, in which equation (1) describes the compliance of the spring as expressed by equation (2).

$$F(X) = kX \tag{1}$$

$$\frac{X}{F(X)} = \frac{1}{k} = \alpha(X) \tag{2}$$

One can plot the relationship of the compliance as a function of displacement, and this plot bears a critical significance when the static compliance (or receptance) is extended over the frequency domain. First of all, we shall assume that a spring and a mass, $m$, are subjected to a harmonic force, which will lead to a harmonic response as expressed by equation (3), where $\omega$ is called the drive frequency. The transfer functions of the spring and the mass are expressed by equation (4). The transfer function of the spring itself does not depend on the drive frequency. The transfer function of the mass depends on the inverse of the squared of the drive frequency. When the two transfer functions are plotted in the Bode diagram (log-log scale), these will be two straight lines forming the skeleton of the *FRF* [6]. The intercept of the two straight lines occurs at the undamped natural frequency of the sprung-mass system. By adding damping, the skeleton can be dressed using the equation (5), where $c$, is the viscous damping and $A$, at the numerator, is an arbitrary real value constant.

$$x(t) = X e^{i\omega t}$$
$$f(X, t) = F(X) e^{i\omega t} \tag{3}$$

$$\frac{X}{F(X)} = \frac{1}{k}$$
$$\frac{X}{F(X)} = -\frac{1}{m\omega^2} \tag{4}$$

$$\frac{X}{F}(\omega) = \frac{A}{k - m\omega^2 + i\omega c} \tag{5}$$

Figure 1 shows a 3D plot of the Bode diagram, where the static receptance is constant because the stiffness function is constant, and the *FRF* will repeat itself for any displacement amplitude of the vibration, as expected for a linear system. The waterfall of linear *FRFs* generates a *linear frequency response* surface that will return a linear *FRF* regardless of the amplitude of the force plane used for cutting across that surface. A constant amplitude force plane extending over the frequency axis (see Figure 1) will produce a linear *FRF*.

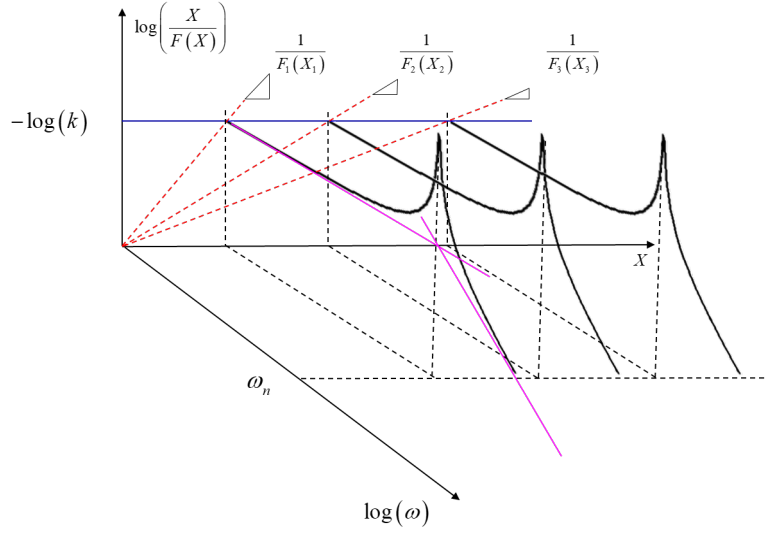

**Figure 1 Expansion of static compliance as a function of displacement on the frequency domain.**

## 2.2 Force-displacement relationship for nonlinear system

In this section, the linear equation of motion expressed by the transfer function in equation (5) is rewritten by including a cubic stiffness nonlinearity, see equation (6). The relationship between force and displacement is nonlinear, for which $k_{LIN}$ is the linear stiffness coefficient and $k_{NL}$ is the nonlinear one. Assuming both harmonic force and response, as given in equation (3), one calculates the static transfer function for $\omega = 0$. Equation (7) shows the static receptance for linear (a) and nonlinear responses (b). These two static transfer functions are plotted in Figure 2.

The next step is to repeat the procedure as described earlier. A cascade of linear *FRFs* will be generated using equation (9) starting from the nonlinear static transfer function $\alpha(X)$.

$$m\ddot{x} + c\dot{x} + k_{LIN}x + k_{NL}x^3 = f(x,t) \tag{6}$$

$$\frac{X}{F}(\omega = 0, X) = \alpha(X) = \frac{1}{k_{LIN}} \quad \text{(a)}$$

$$\frac{X}{F}(\omega = 0, X) = \alpha(X) = \frac{1}{k_{LIN} + k_{NL}X^2} \quad \text{(b)} \tag{7}$$

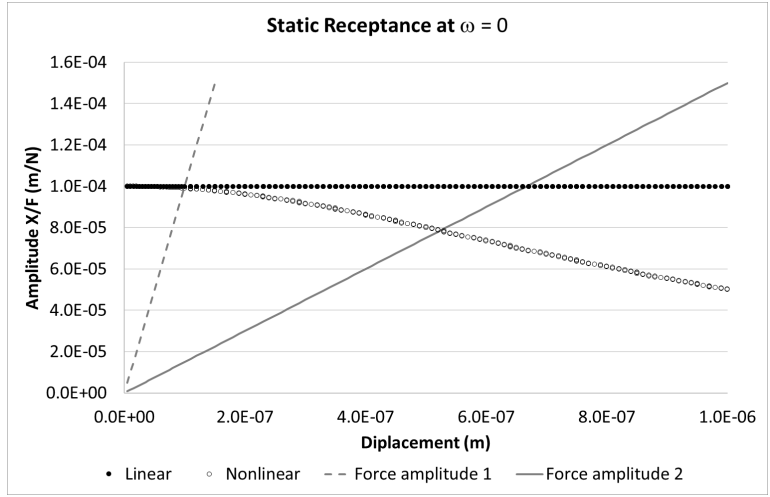

**Figure 2 Linear and nonlinear static receptance curves.**

Figure 2 shows that every point of the nonlinear static receptance is the starting point of a linear *FRF*. The effective stiffness, $K_E$, can be calculated by the equation (8) and used for generating linear *FRFs* by equation (9). A viscous or hysteretic damping model can simulate the *FRF* surface, recalling the relationship between the damping loss factor and the damping ratio $\eta = 2\zeta$.

$$K_E(X) = \frac{X}{F_i(X)} \tag{8}$$
$$i = 1...n$$

$$\alpha(\omega, X) = \frac{{}_rA}{k_E(X) - m\omega^2 + i\omega c} \tag{9}$$

Figure 3 shows a *nonlinear frequency response* surface generated between 0 and 100 Hz in a displacement range between 0 and $1*10^{-6}$ m in the colour spectrum. The surface is built with many linear *FRFs* which follow the pattern set by the nonlinear static receptance. Figure 2 shows two forces of different amplitudes, and the equation (10) allows for evaluating the receptance amplitudes of the nonlinear *FRF* given a constant force amplitude. When that force is extended over the frequency range, it generates a constant amplitude force plane. This force surface cuts across the *nonlinear frequency response* surface, as shown in Figure 4. A nonlinear *FRF* curve is visible on that force surface, as shown in Figure 5.

$$\bar{\alpha}(X) = \frac{X}{F_i} \tag{10}$$

$$F_i(\omega) = \frac{X}{\bar{\alpha}(\omega, X)} \tag{11}$$

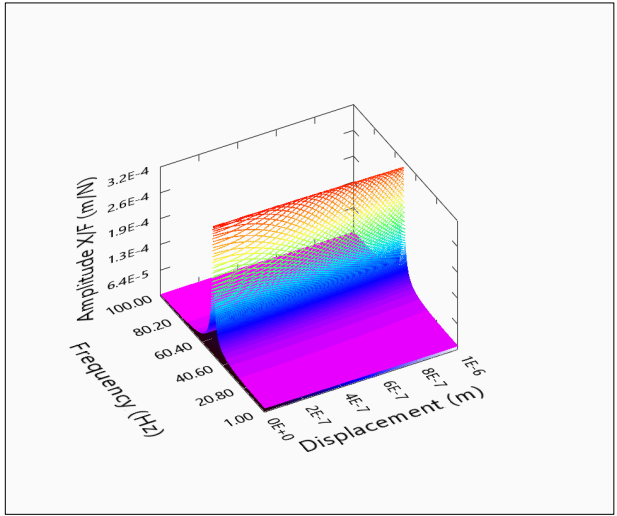
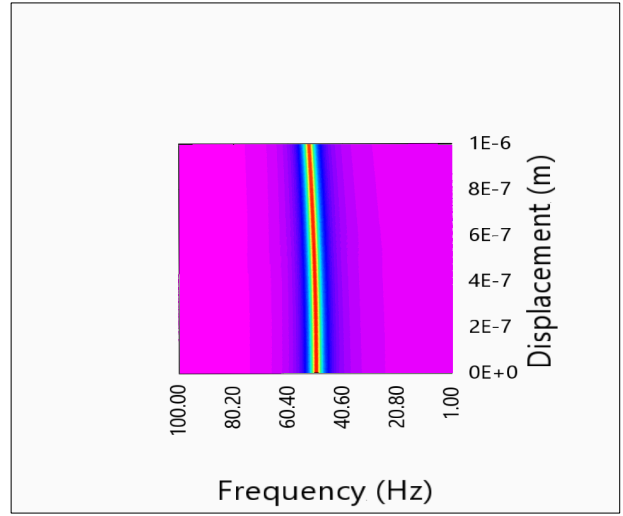

a) Perspective.   b) Top-view.

**Figure 3** *Nonlinear frequency response* **surface**

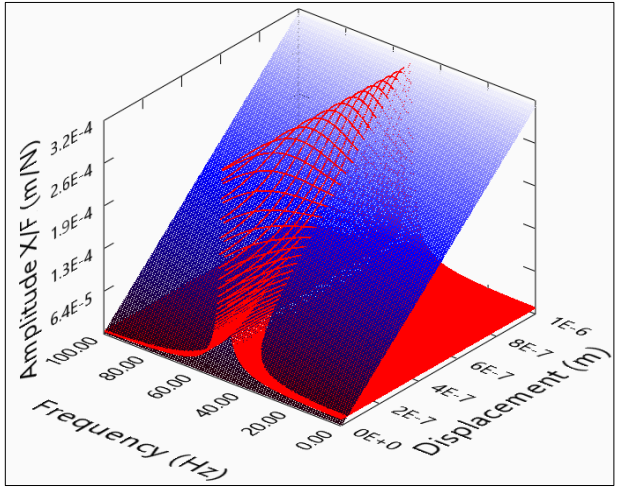
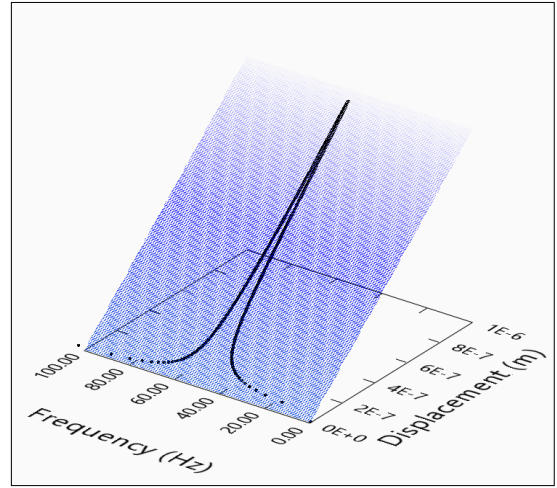

**Figure 4** *Nonlinear frequency response* **surface (red) and constant force plane (blue).**

**Figure 5** Nonlinear *FRF* overlaid on the force surface.

The final step is to extract for a given force the nonlinear *FRF* using the equations (12) and (13). The unknown term is the drive frequency called, $\omega_{NL}$, which is the new drive frequency of the nonlinear *FRF*, as opposed to $\omega$ of the linear *FRF*. The receptance amplitudes are known at every displacement amplitude (see the plot in Figure 2), and the effective stiffness is calculated from equation (7), whereas the mass, $m$, and the damping, $c$, are taken from the Table 1. Reworking the equation (12) as function of the drive frequency, one can solve an equation of the fourth order, equation (13), which will give four roots for the selected receptance value. Once the nonlinear drive frequency, $\omega_{NL}$, is calculated for every linear *FRF* at a given receptance amplitude, the real and imaginary parts of the nonlinear receptance can be calculated using equations (14) and (15). The undamped natural frequency, $\omega_r$, calculated by the ratio between the effective stiffness, $K_E(X)$, at displacement ($X$), and the mass $m$. One will be able to generate the full nonlinear *FRF* for any desired force amplitude, as shown in Figure 6.

$$\left|\bar{\alpha}(\omega, X)_{F_i}\right| = \left|\frac{1}{k_E(X) - m\omega_{NL}^2 + i\omega_{NL}c}\right| \quad (12)$$

$$\omega_{NL}^4 + \left(2mk_E(X) - \frac{cmk_E(X)}{\sqrt{mk_E(X)}}\right)\omega_{NL}^2 + \frac{k_E^2(X)}{m^2} - \frac{1}{|\bar{\alpha}(\omega,X)_{F_i}|} = 0 \quad (13)$$

$$RE_{1,2} = \omega_r^2 - \omega_{NL1,2}^2 \quad (14)$$

$$IM_{1,2} = 2\zeta * \omega_r * \omega_{NL1,2} \quad (15)$$

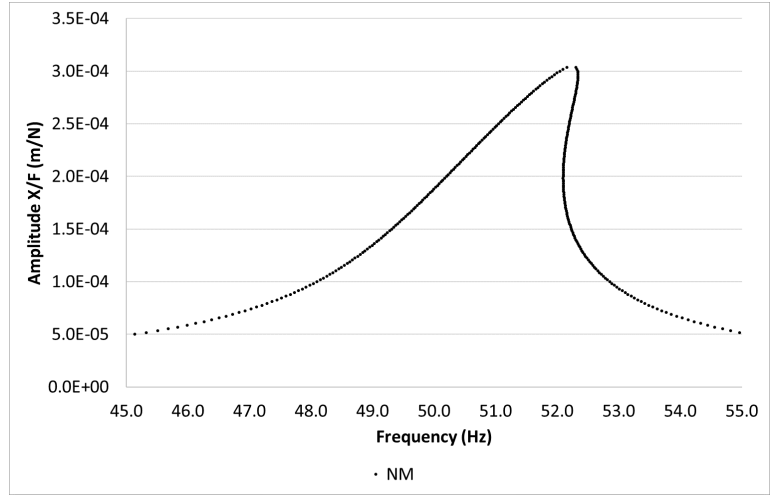

**Figure 6 Nonlinear *FRF* calculated from the *nonlinear frequency response* surface.**

The verification of the nonlinear *FRF*, thus calculated, is carried out by numerical integration (ODE) and the Harmonic Balance Method (HBM). A single DOF equation of motion (16) generates nonlinear *FRFs*. The system parameters implemented in the equation (16) are listed in the Table 1. The simulation of the nonlinear *FRF* was run between 40 and 60 Hz with a frequency step of 0.1 Hz, and every step was simulated for 2 seconds at 10,000 samples/sec. The HBM is performed using the first fundamental harmonic, as many textbooks show. Instead of the modulus and phase, the real and imaginary parts of the *FRF* are calculated to plot the Nyquist circles. Figure 7 shows Nyquist curves for the three calculated *FRFs*, which overlay on each other.

$$\ddot{x} + 2\zeta\omega_r\dot{x} + \omega_r^2 x + k_{NL}x^3 = p = \frac{f(t)}{m} = P\cos(\Omega t + \phi) \quad (16)$$

| $m$ = 0.1 kg | $c$ = 1 m/s N$^{-1}$ | $k_{LIN}$ = $10^4$ N/m | $k_{NL}$ = $1^{15}$ N/m³ |
|---|---|---|---|

**Table 1 SDOF system parameters**

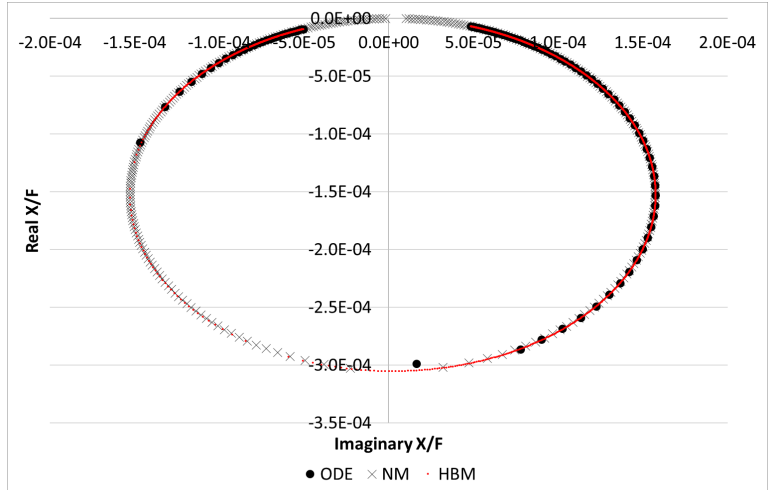

**Figure 7 Comparison of the Nyquist circles simulated by numerical integration (ODE), the new method (NM) and the harmonic balance method (HBM).**

The single Degree of Freedom model demonstrated that a nonlinear transfer function can be evaluated at a given excitation force from a *nonlinear frequency response* surface. These three *FRFs* overlay perfectly on top of each other, thus validating the hypothesis. The new analysis method postulates that the unknown parameters are the nonlinear drive frequencies and unique solutions in the force plane. The nonlinear vibration amplitudes do not coexist at the same drive frequency, as seen in an *FRF* plotted in a two-dimensional plane, but they belong to different nonlinear drive frequencies. The following section will expand the new theoretical formulation to a TWO-DoF system with a grounded cubic nonlinearity.

## 2.3 The theoretical nonlinear transfer function for a TWO-DoF system

This section will investigate the same theoretical approach using a TWO-DoF system featuring a cubic nonlinearity. The static transfer function will generate a surface, and, as done before, a force plane will be used to calculate a nonlinear transfer function, which will be verified through an *FRF* simulated by numerical integration.

## 2.4 Simulation of nonlinear *FRFs* from a TWO-DoF system

This subsection extends the same theoretical formulation to a multi-degree-of-freedom system with a grounded cubic nonlinearity, as shown in Figure 8. The system is made of two masses, $m$, and three equal linear springs, $k_{LIN}$, one of the two grounded springs will exhibit nonlinear behaviour, as shown in Figure 8. The equations of motion are written for the undamped system and presented in equation (17). Table 2 reports the system's and simulations' parameters for the steady state analysis using numerical integration.

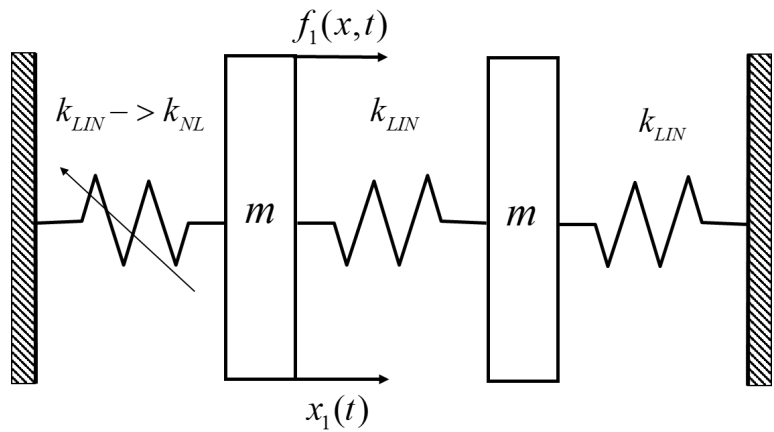

**Figure 8 TWO-DoF system with grounded nonlinearity.**

| Mass (kg) | Stiffness (N/m) | Damping Ns/m | Nonlinear stiffness (N/m^3) | Excitation force (N) | Frequency step (Hz) | Sample rate Sample/sec | Time generation (sec) | Steady-state part (sec) |
|---|---|---|---|---|---|---|---|---|
| 0.1 | 15,000 | 1 | $10^{14}$ | 0.00124 | 0.05 | 10,000 | 2 | 0.3 |

**Table 2 System parameters and measurement parameters for the steady state simulations.**

Under steady state conditions, one can assume a harmonic excitation and response, as expressed in equation (3), and then solve for the transfer function at the first degree of freedom $x_1(t)$, as expressed in equation (18). At this stage, it is unnecessary to include any damping in the equations of motion, and the inertia terms will drop off because the static transfer function is evaluated at a null frequency.

$$\begin{cases} m\ddot{x}_1 + k_{LIN}x_1 + k_{NL}x_1^3 + k_{LIN}(x_1 - x_2) = f_1(x,t) \\ m\ddot{x}_2 - k_{LIN}(x_1 - x_2) + k_{LIN}x_2 = 0 \end{cases} \quad (17)$$

To avoid confusion with equation (10), the force at the DoF $X_1$ is written as follows $F^{(1)}$.

$$\frac{X_1}{F^{(1)}}(\omega, X) = \frac{2k - m\omega^2}{-m\omega^2(2k - m\omega^2) + k(2k - m\omega^2) + k_{NL}(2k - m\omega^2)X_1^2\cos^2(\omega t) + k(2k - m\omega^2) - k^2} \quad (18)^1$$

The drive frequency is set to zero to evaluate the static transfer function, as done for the single DOF system. Equation (19) allows us to calculate the equivalent nonlinear stiffness as a displacement function. When the nonlinearity is null, the static transfer function returns the same value for all displacements, as expected for linear vibrations.

$$\frac{X_1}{F_i^{(1)}(X)} = K_E(X) = \frac{2k_{LIN}}{2k_{LIN}k_{NL}X_1^2 + 3k_{LIN}^2} \quad (19)$$

A standard transfer function for a viscous model generates linear *FRFs*, forming the *nonlinear frequency response* surface. Equation (18) calculates the linear frequency responses, where the effective stiffness is calculated from equation (19).

$$\alpha_{1,1}(\omega, X) = \left([K(X)] - \omega^2[M] + i\omega[C]\right)^{-1}$$

$$[K(X)] = \begin{bmatrix} 2k_E(X) & -k_E(X) \\ -k_E(X) & 2k_E(X) \end{bmatrix} \quad (20)$$

Figure 9 shows the *nonlinear frequency response* surface generated by the equation (20), and Figure 10 shows a top view of that figure where the variation of the resonance frequency of the first and second modes vary as a function of the displacement ($X_1$).

The process to evaluate the nonlinear *FRF* repeats, as explained earlier, Figure 11 shows the surface (red) cut across by an arbitrary force plane (blue). Extracting the nonlinear *FRF* from the surface would require solving the equation (20) where the drive frequency, $\omega_{NL}$, is the unknown parameter and the amplitude of the static transfer function is known at every displacement amplitude. The analytical solution for the nonlinear drive frequency can be very cumbersome, so a much simpler solution technique is offered.

---

[1] $k_{LIN} = k$

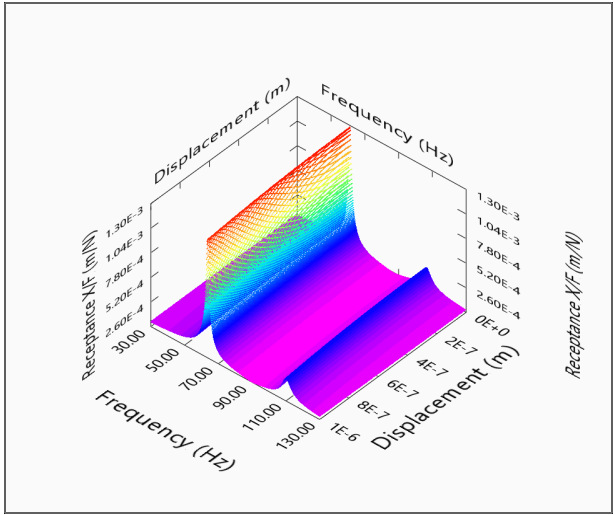 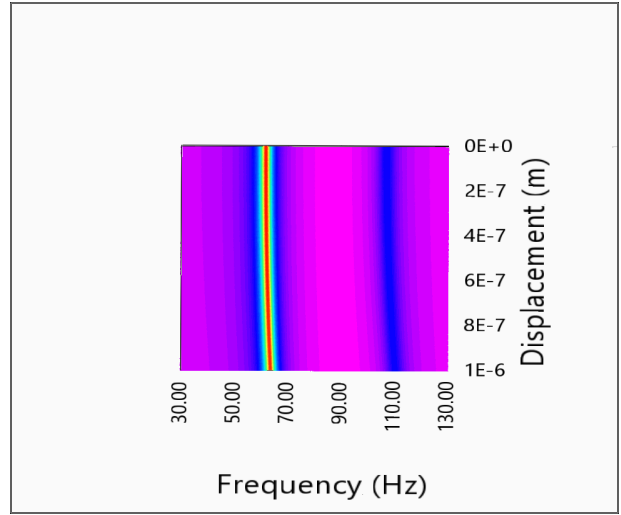

**Figure 9** *Nonlinear frequency response* **surface generated by equation (20).**

**Figure 10 Top-view of the *Nonlinear frequency response* surface.**

The product shown in equation (21) equals unity for frequencies evaluated at the same given transfer function $\bar{\alpha}(\omega, X)_{F_i}$; otherwise, it is different. A numerical solution is used for calculating the new nonlinear drive frequency, $\omega_{NL}$, for which a search condition is implemented. The finer the frequency steps forming the linear *FRF* of the *nonlinear frequency response* surface, the smaller the error for the search condition that can be set.

$$\left| \bar{\alpha}(\omega, X)_{F_i} \right| \left\| \left([K(X)] - \omega_{NL}^2 [M] + i\omega_{NL}[C]\right) \right\| = 1 \qquad (21)$$

The first verification is done by setting the nonlinear coefficient to null, extracting a linear *FRF* compared to one generated by numerical integration. The numerical integration was somewhat redundant, but consistency was maintained with the simulation when the nonlinearity was activated again. Figure 12 shows the comparison and the perfect overlay between the two *FRFs*.

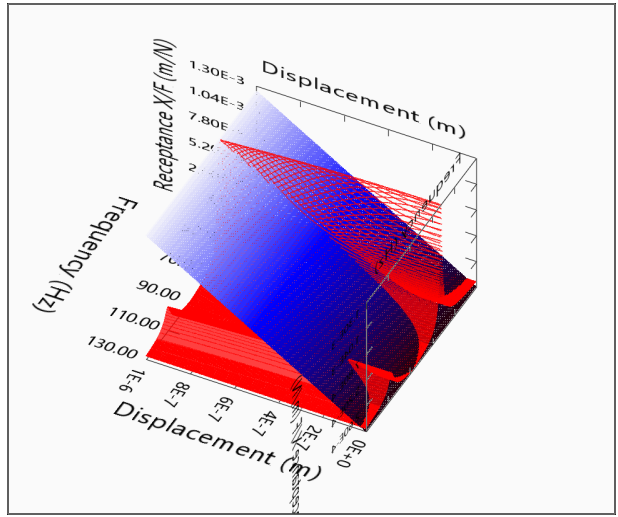

**Figure 11** *Nonlinear frequency response* **surface (red) and a constant force plane (blue) for the extraction of the nonlinear *FRF*.**

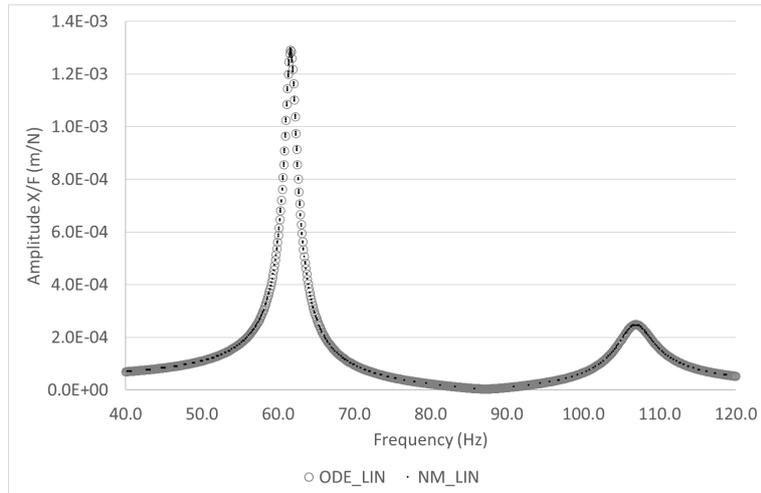

**Figure 12 Comparison between linear *FRF* using numerical integration and NM.**

The second simulation is yielded by activating the nonlinearity, setting the nonlinear stiffness coefficient of the cubic term to the value listed in Table 2. The *FRF* extracted from the surface is overlaid by the one generated by the numerical integration, as shown in Figure 13. Both the amplitude and phase of the *FRFs* overlay well, for which the first mode shows a more distinct nonlinearity than a much milder one of the second mode.

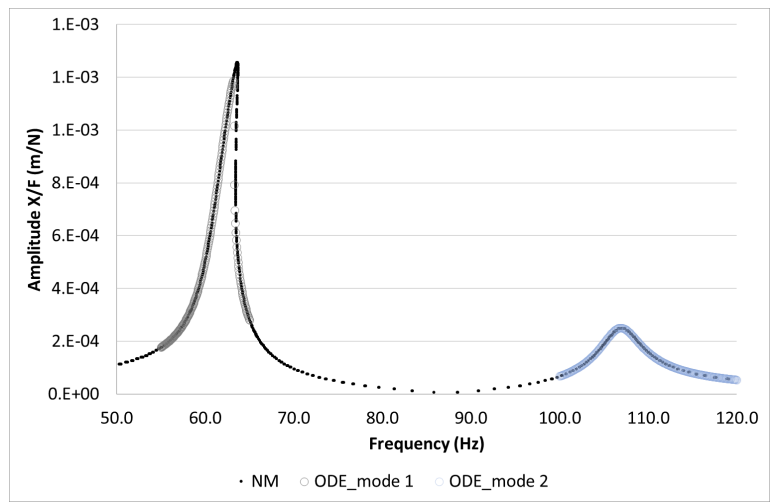

(a) nonlinear *FRF* modulus

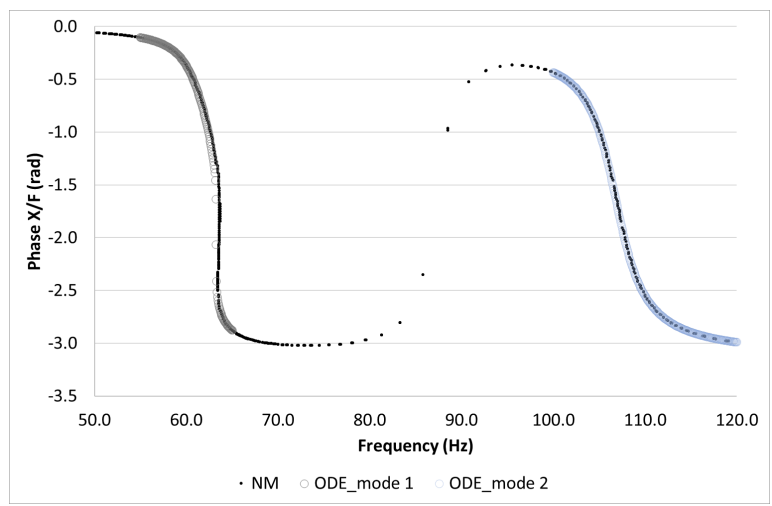

(b) nonlinear *FRF* phase

**Figure 13 Nonlinear *FRF* generated by the new analysis method and numerical integration.**

## 2.5 Final remarks

This section proved the hypothesis formulated in the introduction using both ONE- and TWO-DoF systems with a cubic nonlinearity. The *nonlinear frequency response* surface can be generated when a static nonlinear transfer function is available. That function sets the equivalent stiffness for every linear *FRF*, which will build a nonlinear surface. The denser the displacement steps, the smoother the surface becomes. When an arbitrary constant amplitude force plane cuts across that surface, the nonlinear drive frequencies can be calculated and used to generate a nonlinear *FRF*. The new analysis method allows for identifying unique drive frequencies on a three-dimensional surface.

In the first case, simple mass-spring single and multi-DoF systems calculated a static nonlinear transfer function. In both cases, a force was applied to deflect the mass-spring systems. That approach worked because the nonlinearity was grounded, and the transfer function could be derived for one degree of freedom, such as $X_1$. However, there are situations where nonlinearity is not grounded, but it appears between the two degrees of freedom. When that happens, it is clear that static compliance remains a function of both degrees of freedom, such as $\frac{X_1}{F_1}(\omega = 0) = \alpha(\omega = 0, X_1, X_2)$. The problem can be solved similarly to that of Singh et al.[7] described in their papers, but without using the mode shapes. It is possible to use an arbitrary deflection shape, such as $X_1 = -X_2$, which uncouples the equations and allows the calculation of static nonlinear compliance. Using operational deflections rather than mode shapes avoids the modal analysis and keeps the whole approach bound to the response space. Magi et al. applied the deflection shape method.[8] and Di Maio et al.[9] for calculating the energy release rate (ERR) of fatigue damage growth. Section 2 did not include any nonlinear damping because the intention was to evaluate a new analysis method for synthesising a nonlinear *FRF* assuming a cubic stiffness nonlinearity. Furthermore, nonlinear damping will be observed in the experimental verification section. Finally, two examples in Appendix A2 show how the theoretical formulation can be exploited for *FE* analyses.

The next challenge is to prove that the same analysis method works when the input parameter is not a nonlinear static receptance but the input is the amplitude-dependent modal parameters. The following section will propose a novel method for identifying nonlinear modal parameters.

## 3 Modal analysis of nonlinear response functions

This section will present some background literature about identification methods of nonlinear modal parameters that can be extracted by *FRF* analysis using the single input, single output (SISO) testing method. It will then discuss how the Dobson formulation for linear system identification can be modified to calculate amplitude-dependent modal parameters.

### 3.1 Background literature

Measuring the frequency response functions is the most used experimental technique for characterising smooth nonlinearities. Although time-consuming because it requires steady state conditions, the stepped-sine test is convenient because it deals with sine IN - sine OUT (+ noise + nonlinear harmonics). The ratio between the complex Fourier terms of response and the force at the drive frequency generates the frequency response. With the inclusion of a PID controller, able to control the amplitude of the steady state of the response for every drive frequency, the *FRFs* can be "*linearised*" at a given constant response amplitude, and linear modal analysis methods can process the transfer functions. The higher the amplitude levels, the more the nonlinearity can be characterised [10-12]. In the nineties, Li [13] investigated a different technique for analysing nonlinear *FRFs* in his PhD thes*is*. It did not require a control method for acquiring vibrations but was based on the open-loop control technique. The identification method was based on analysing a nonlinear *FRF* by taking two pairs of frequency points at equal vibration amplitude on either side of the response peak. Hence, one could write two receptance *FRF* equations at the same vibration amplitude, one for the frequency at the left and one at the right of the maximum response peak. These two equations, Real and Imaginary relationships, could be used to calculate the modal parameters. By sweeping the pair of frequency points from low up to high vibration amplitude, the nonlinear *FRF* could be easily characterised.

Carrella and Ewins [14] investigated this technique using an aerospace structure and evaluated its weakness. The method fails when one *FRF* branch is absent due to the system's unstable vibration response. Hence, the interpolation method used to identify the frequency points at the same displacement amplitude cannot be applied, so the equations cannot be solved correctly. Furthermore, the analysis is based on the single degree of freedom theory, which does not consider the contribution of neighbour modes. Although simple and practical, this experimental technique presents many limitations.

In recent years, Zhang and Zang [15, 16] yielded several research papers presenting a method for identifying nonlinear modal parameters by measuring and analysing *FRFs*. Every *FRF* is measured by step-sine tests under steady state vibrations and labelled at drive voltage (V) feeding the shaker. The very innovative idea is to generate a response surface (such as acceleration) made of frequencies, voltages and accelerations [Hz, V, m/s^2] and a force surface made of frequencies, voltages and forces [Hz, V, N]. The linearisation process is achieved by extracting a constant acceleration curve from the acceleration surface and using it to extract the force curve. This operation is achieved using the common voltage axis to identify accelerations and forces. Once the operation is completed, the *FRF* linearised at the constant acceleration is derived. Finally, the linear modal analysis can extract the modal parameters. The more *FRFs* extracted from the surface, the better the nonlinearity characterisation is. The method is very robust for smooth and non-smooth types of nonlinearities. These techniques can calculate the amplitude-varying modal parameters from linear *FRFs*. The following subsection will propose a new technique derived from the line-fit method using the Dobson formulation. Before doing that, some basics of modal analysis using the inverse *FRF* methods are provided. The assumption of well-separated modes is still valid in this paper section.

## 3.2   Inverse methods for modal analysis

The inverse *FRF* methods are commonly used for single DoF modal analysis[1]. The convenience of the analysis is that the inverse of an *FRF* can be separated into the Real and Imaginary parts and plotted as a function of the drive frequency. Around the resonance, these two functions are straight lines, as becomes clear by inspecting the equation (22), where $\omega_r$ is the natural frequency, $\eta_r$ is the damping loss factor of the $r$ mode shape.

$$\alpha(\omega) = \frac{_rA}{\omega_r^2 - \omega^2 + i\zeta_r \omega \omega_r} = \frac{_rA}{\omega_r^2 - \omega^2 + i\eta_r \omega_r^2} \qquad (22)$$

The modal constant $_rA$ in equation (22) can be either a real or complex number. When the constant is a real number, equation (22) can be inverted and separated into its real and imaginary parts, which can be plotted as a function of the squared drive frequency, $\omega^2$. The real part of equation (23) is used for calculating the natural frequency, while the imaginary part of the same equation is used for calculating the damping, either the loss factor or the damping ratio.

$$\alpha(\omega)^{-1} = RE + IM = \frac{\omega_r^2 - \omega^2}{_rA} + i\frac{\eta_r \omega_r^2}{_rA} = \frac{\omega_r^2 - \omega^2}{_rA} + i\frac{\zeta_r \omega \omega_r}{_rA} \qquad (23)$$

When the constant $_rA = A_r + iB_r$ is a complex number, the inverse can be calculated by multiplying the numerator and denominator by its complex and conjugate, thus leading to (24). From this point, the manuscript will refer to a hysteretic damping model coherent with the Dobson formulation discussed hereafter.

$$\alpha(\omega)^{-1} = \frac{\omega_r^2 - \omega^2 + i\eta_r \omega_r^2}{_rA} = \frac{(A_r + B_r\eta_r)\omega_r^2 - A_r\omega^2}{A_r^2 + B_r^2} + i\frac{(A_r\eta_r + B_r)\omega_r^2 - B_r\omega^2}{A_r^2 + B_r^2} \qquad (24)$$

The real and imaginary parts of the equation (24) are linear relationships as a function of $\omega^2$, see equation (25). These two equations (25a&b) can be solved by calculating the coefficients and intercepts of the straight lines, which are used for evaluating the four modal properties $\left(\omega_r^2 \quad \eta_r \quad A_r \quad B_r\right)$.

$$RE\left(\alpha(\omega)^{-1}\right) = \frac{(A_r + B_r\eta_r)\omega_r^2 - A_r\omega^2}{A_r^2 + B_r^2} = m_R + n_R\omega^2 \quad (a)$$

$$IM\left(\alpha(\omega)^{-1}\right) = \frac{(A_r\eta_r + B_r)\omega_r^2 - B_r\omega^2}{A_r^2 + B_r^2} = m_I + n_I\omega^2 \quad (b)$$

(25)

Although very simple and intuitive, this method is limited by the upper and lower residuals, which are not eliminated. The method works for well-spaced resonances but fails as soon as resonances become closer in frequency. As opposed to the simple line-fit, Dobson [5] developed a method based on better mathematical formulation, which could eliminate the effect of the residuals from the analysis. Even though the Dobson method is applied to SISO tests, Maia extended the same method to Single-Input Multiple-Output (SIMO) tests [17]. Recalling the drive-point *FRF* in equation (26), the linear frequency response is the total sum of the contribution of $N$ modes ($r$), or equally, the sum of a single mode ($r$) with a constant residual.

$$\alpha_{ii}(\omega) = \sum_{r=1}^{N} \frac{{}_rA_{ii}}{\omega_r^2 - \omega^2 + i\eta_r\omega_r^2} = \frac{{}_rA_{ii}}{\omega_r^2 - \omega^2 + i\eta_r\omega_r^2} + residual \quad (26)$$

The Dobson method introduces a new variable, a pseudo drive frequency, $\Omega$, which is used to build a pseudo response frequency, $\alpha(\Omega)$, near the resonance. The effect of residuals on the response of the mode ($r$) is eliminated by subtracting the two transfer functions, as presented in equation (27).

$$\alpha_{ii}(\omega) - \alpha_{ii}(\Omega) = \frac{A_r + iB_r}{\omega_r^2 - \omega^2 + i\eta_r\omega_r^2} - \frac{A_r + iB_r}{\omega_r^2 - \Omega^2 + i\eta_r\omega_r^2} =$$

$$A_r + iB_r \left[\frac{\omega^2 - \Omega^2}{(\omega_r^2 - \omega^2)(\omega_r^2 - \Omega^2) - \eta_r^4\omega_r^4 + i\eta_r\omega_r^2(2\omega_r^2 - \omega^2 - \Omega^2)}\right]$$

(27)

By multiplying the numerator of equation (27) by the complex and conjugate of the modal constant and by taking the inverse, one can define a function delta $\Delta$, as expressed in equation (28).

$$\Delta = \frac{\omega^2 - \Omega^2}{\alpha_{ii}(\omega) - \alpha_{ii}(\Omega)} = \frac{A_r - iB_r}{A_r^2 + B_r^2}\left(\omega_r^2 - \omega^2\right)\left(\omega_r^2 - \Omega^2\right) - \eta_r^4\omega_r^4 + i\eta_r\omega_r^2\left(2\omega_r^2 - \omega^2 - \Omega^2\right) \quad (28)$$

The "fixing" frequency, $\Omega$, will sweep the entire drive frequency vector $\omega$, thus generating an array $[\Delta]_{n\times n}$ where the zeros obtained for $\Omega = \omega$ will need to be eliminated. The next step is to separate the real and imaginary deltas and plot them as a function of the frequency squared, $\omega^2$, in equation (28).

$$RE(\Delta) = m_R\omega^2 + c_R$$
$$IM(\Delta) = m_I\omega^2 + c_I$$

(29)

The angular coefficients are expressed in equation (30).

$$m_R = -\frac{1}{A_r^2 + B_r^2}\left[A_r\left(\omega_r^2 - \Omega^2\right) + B_r\eta_r\omega_r^2\right]$$

$$m_I = -\frac{1}{A_r^2 + B_r^2}\left[B_r\left(\omega_r^2 - \Omega^2\right) - A_r\eta_r\omega_r^2\right]$$

(30)

The angular coefficients from the plots (RE($\Delta$), $\omega^2$) and (IM($\Delta$), $\omega^2$) are selected and used for generating two new straight lines, as expressed in equation (31).

$$m_R = n_R \Omega^2 + d_R$$
$$m_I = n_I \Omega^2 + d_I \quad (31)$$

One can solve the modal parameters using the relationships expressed in equation (30) by evaluating the two angular coefficients and intercepts of the two straight lines.

$$n_R = -\frac{(A_r + B_r \eta_r)\omega_r^2}{A_r^2 + B_r^2}$$

$$d_R = \frac{A_r}{A_r^2 + B_r^2}$$

$$n_I = -\frac{(A_r \eta_r - B_r)\omega_r^2}{A_r^2 + B_r^2} \quad (32)$$

$$d_I = -\frac{B_r}{A_r^2 + B_r^2}$$

The four constants evaluated by equation (32) can be used for extracting the four modal parameters $(\omega_r^2 \;\; \eta_r \;\; A_r \;\; B_r)$. The Dobson method proved robust because the procedure performs the line fitting twice before the modal parameters are extracted.

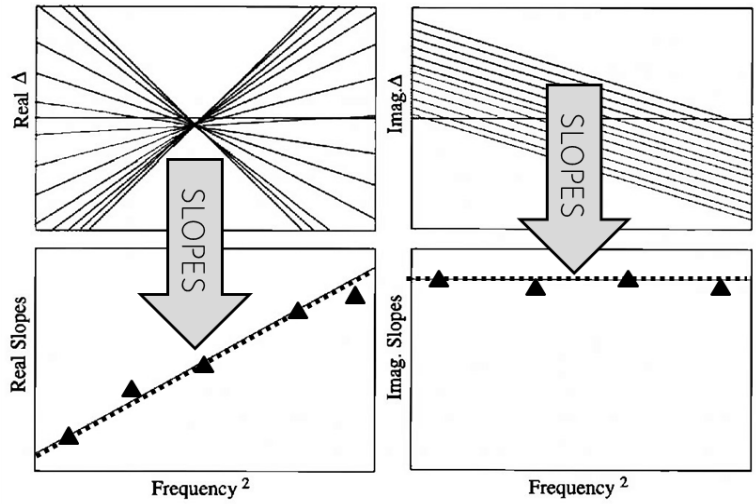

**Figure 14 Implementation of the Dobson method for a linear *FRF*. The slopes from the Real(Δ) and Imag(Δ) are selected and fitted again by straight lines.**

## 3.3  Modified-Dobson method applied to nonlinear *FRFs*

The previous section presented the Dobson method applied to a linear *FRF*. The same mathematical formulation can characterise nonlinear *FRF* and calculate the nonlinear modal parameters. One shall think of a nonlinear *FRF* that starts from an underlying linear response, becoming increasingly nonlinear as one sweeps the Nyquist circle and then more and more linear as one approaches its origin again. The Nyquist points do not indeed form a circle, but it is possible to select three frequency points forming a Nyquist circle of a linear system. Assume that one takes two frequency points from the lowest amplitude of the *FRF* at which the vibration response can be considered linear.

Moreover, these two points can be taken on either side of the max response peak. Now, one shall assume that a third point is taken at any desired amplitude of the receptance, thus forming a triplet of frequency points. This triplet can form a Nyquist circle, identifying an equivalent linear system for the selected receptance amplitude. By sweeping the third frequency from the first to the second reference point of the *FRF*, one can observe how the modal parameters change from linear to nonlinear vibrations. In principle, it is like deconstructing a nonlinear *FRF* into a series of linear ones referenced at different amplitudes.

The data analysis is therefore carried out by selecting triplets of frequency responses at every iteration. These triplets are selected as indicated in the previous paragraph. The two references are called the "*fixers*", and these two will never change during the Dobson analysis. The third frequency point, forming the triplet, is called the "*sweeper*" because it will change for every triplet. Therefore, a selected response peak of an *FRF* made of 32 frequency points will be analysed 30 times, such that three frequency points are taken for each analysis, as expressed in equation (33). The *sweeper* will sweep the frequency points between the two references.

$$\left(\alpha\left(\omega_1^{fxr}\right), \alpha\left(\omega_i^{swp}\right), \alpha\left(\omega_N^{fxr}\right)\right)$$
$$i = 2,\ldots,N-1$$
(33)

Therefore, the procedure is repeated from equation (26) up to (32) as often as needed for the *sweeper* to observe all the frequency points around the response peak. Figure 15 shows a diagram of the modified-Dobson implementation for the analysis of a nonlinear *FRF*, and it becomes more apparent when Figure 16 is also observed.

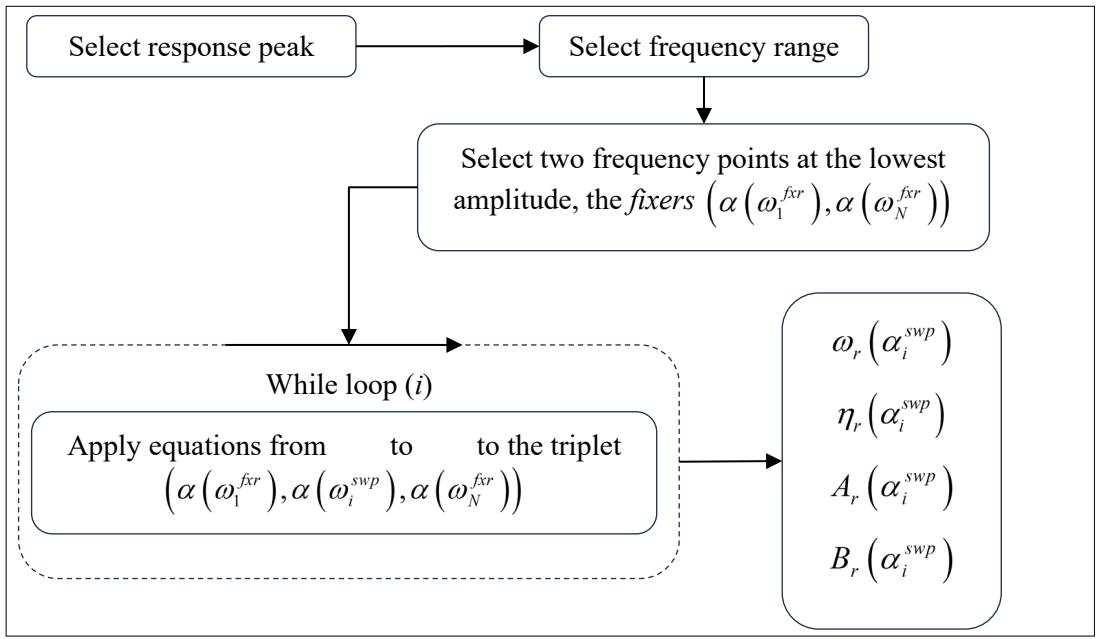

**Figure 15 Diagram of the modified-Dobson method.**

Figure 16a shows the Nyquist circle, while Figure 16b shows the moduli of the *FRF*s. The two frequency points in green are the *fixers*, and these are at low amplitudes where the vibration response is still linear, as visible in Figure 16b. The third frequency point is the *sweeper*, in red, which can be taken anywhere between the two references, and the modal parameters will be a function of the modulus of the receptance $\left|\alpha\left(\omega_i^{swp}\right)\right| = \alpha_i^{swp}$. This triplet can be used to generate the Nyquist circle of an equivalent linear system at the amplitude of the *sweeper*. The Dobson method is now applied to the triplet to extract the modal parameters at that given amplitude, as the diagram in Figure 15 describes.

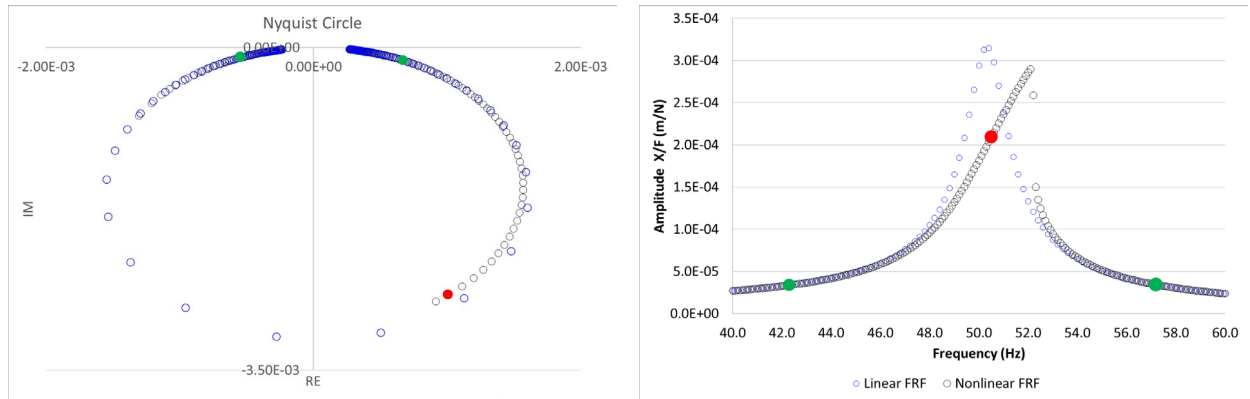

a) Nyquist plots

b) Modulus of linear and nonlinear *FRF*

**Figure 16 Linear and nonlinear response functions, with fixers (green) and sweeper (red).**

Figure 17 shows one step of the straight-line fitting, as expressed in equation (29) and where the coefficients are expressed in equation (30). The natural frequency can be evaluated as a function of the amplitude of the *sweeper* by repeating the process as many times as required. Figure 17a shows the real part of the delta, while Figure 17b shows the imaginary part.

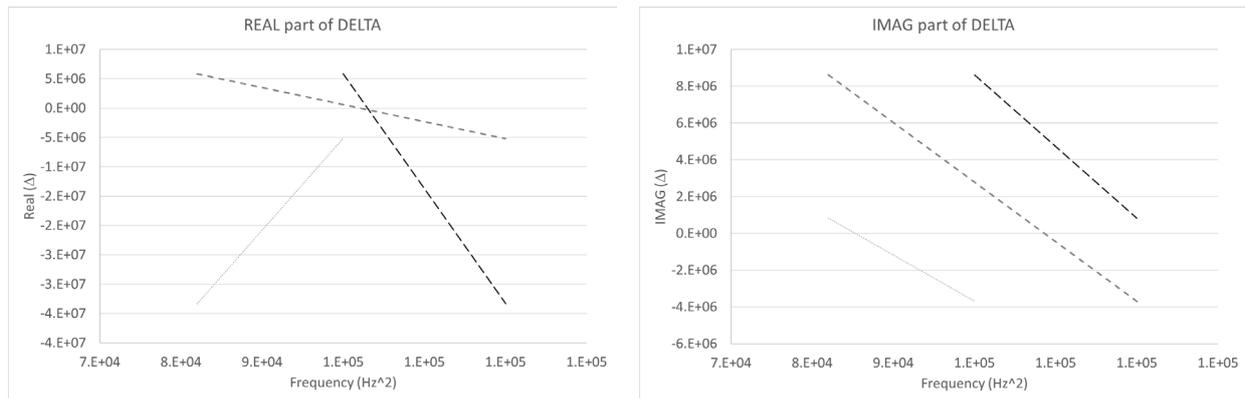

a) Real ($\Delta$) of the first step analysis

b) Imaginary ($\Delta$) of the first step analysis

**Figure 17 The Real and Imaginary $\Delta$ of the first step analysis of one triplet.**

The red dot in Figure 18 indicates the natural frequency of an equivalent linear system at the amplitude of the *sweeper*. Figure 18 also shows the natural frequency curve where some points are circled in blue, indicating the natural frequency curve calculated when the *sweeper* crosses over the response jump, from about 52 Hz up to 53 Hz. The modified-Dobson method is a simple and effective technique that overcomes two main shortfalls identified in the Lin's and the Zhang-Zang methods. The first one is limited when one of the two branches of the *FRF* is missed, as in the example presented in Figure 18, because the algorithm needs two frequency points on either side of the max response peak of the *FRF*. The second method requires the measurement of a few nonlinear *FRFs* to extract linearised ones, which can be used to evaluate the modal parameters. The modified-Dobson method can extract the nonlinear modal parameters from a single *FRF*. The applicability of the modified-Dobson method is valid for well-separated resonances because equation (27) eliminates the residuals from neighbouring modes, but it fails when modes become closer and closer for two main reasons. For the first reason, the two references are taken at an amplitude with linear vibration. However, the closer the modes are, the higher the chances of including nonlinear residuals in frequency points taken as references. In the second one, the *sweeper* sweeps the frequency points from one fixer to the other, thus including different residuals at every iteration. The standard Dobson method equation (25) is calculated for as many frequency points as available in the analysis run, thus eliminating the residuals. In the present modified formulation, equation (27) is carried out for one triplet, which changes every time. The issue can be mitigated by including a technique called "interference criteria in modal identification" proposed by Maia [18].

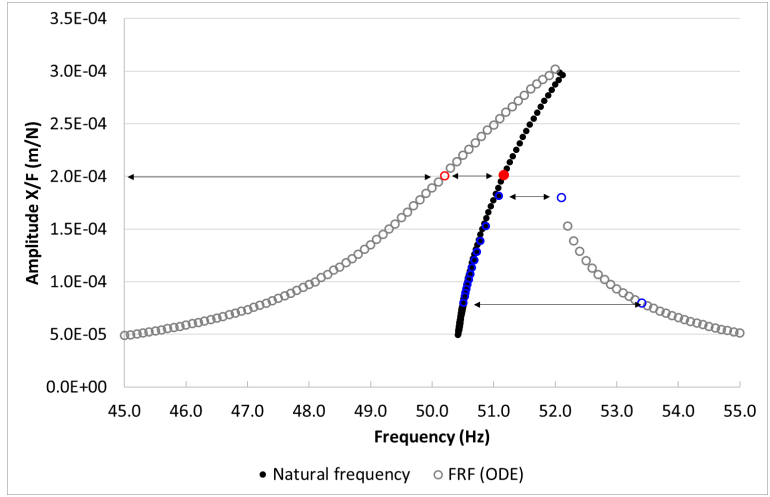

**Figure 18 Nonlinear *FRF* with natural frequency in black. The red dot indicates the linearised natural frequency at the amplitude of the *sweeper*.**

Finally, having evaluated the modal parameters as a function of the amplitude of the *sweeper*, one shall convert those into polynomial functions. Table 3 reports the steps to generate the polynomial functions.

| Function of receptance | Function of displacement | Polynomial functions | |
|---|---|---|---|
| $\omega_r(\alpha_i^{swp})$ | $\omega_r(X_i^{swp})$ | $\omega_r(X) = \omega_n X^n + ... + \omega_1 X + \omega_0$ | (34) |
| $\eta_r(\alpha_i^{swp})$ | $\eta_r(X_i^{swp})$ | $\eta_r(X) = \eta_n X^n + ... + \eta_1 X + \eta_0$ | (35) |
| $A_r(\alpha_i^{swp})$ | $A_r(X_i^{swp})$ | $A_r(X) = A_n X^n + ... + A_1 X + A_0$ | (36) |
| $B_r(\alpha_i^{swp})$ | $B_r(X_i^{swp})$ | $B_r(X) = B_n X^n + ... + B_1 X + B_0$ | (37) |

**Table 3 Summary table for the modal properties.**

Equation (36) can be used to generate the experimental *nonlinear response frequency* surface.

$$\alpha(\omega, X) = \frac{{}_rA(X)}{\omega_r^2(X) - \omega^2 + i\eta_r(X)\omega_r^2(X)} \tag{38}$$

It is worth mentioning that equation (38) is written as a function of displacement. However, it could be the same if the velocity or acceleration were measured, so one will deal with Mobility or Acceleration *FRFs*. The *nonlinear frequency response* surface will be coherent with the units used for the measurements, and conversions are always possible because of the steady state conditions. Equation (39) can be solved numerically, as indicated in section 2.4, to find the nonlinear drive frequencies, $\omega_{NL}$, of the nonlinear *FRF*.

$$\left| \bar{\alpha}(\omega, X)_{F_i} \right| \left\| \left( \frac{{}_rA(X)}{\omega_r^2(X) - \omega_{NL}^2 + i\eta_r(X)\omega_r^2(X)} \right)^{-1} \right\| = 1 \tag{39}$$

The following section will demonstrate that the nonlinear *FRF* evaluated from the synthesised surface can be compared to the experimental one.

## 4  Experimental validation

This section will describe the experimental validation of the modified-Dobson method and the application of the new method for extracting nonlinear *FRFs*. The validation is developed in three stages. The first experimental validation (setup-1) will verify that the proposed modified-Dobson method produces the same results as the Zhang-Zang method. The second and third experimental validations will verify that the new analysis method can calculate nonlinear *FRFs* once the nonlinear modal parameters are identified. These parameters are used to regenerate *FRFs* by the equation (38) to build the *nonlinear frequency response* surface, from which the nonlinear *FRFs* are calculated.

As mentioned earlier, equation (36) is a receptance model that converts to Mobility or Acceleration depending on the test data. Three test setups, built over the past ten years for different research reasons and applications, are used for this validation exercise.

## 4.1 Validation of the modified-Dobson method using setup-1

The experiment for validating the modified-Dobson method is made of a simple aluminium semicircle structure with two extremes connected by two strips of metals. The test structure is mounted by a load cell to the shaker head, while an accelerometer is connected to the side of the semicircle. Figure 19 shows the test structure and the setup with the shaker and sensors. The mode of vibration selected for this experiment is at approximately 326 Hz, which brings the two extreme points of the arch close and apart, thus forcing the thin metal strips into nonlinear vibrations.

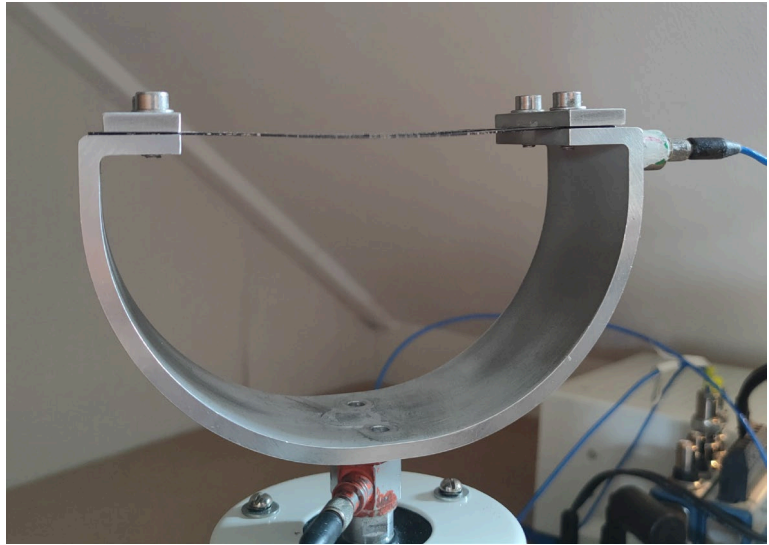

**Figure 19 Test structure for validation of the modified-Dobson method.**

A custom-made acquisition panel was developed to measure the acceleration and force signals and calculate the transfer functions. The control panel stepped over the frequency range of interest using a drive signal for the shaker, shaped with a cosine tapered window. Therefore, every sinewave had a smooth ramp-up, a steady-state part and a smooth ramp-down signal. It avoided harsh transients from one frequency to the next one. Figure 38, in Appendix A3, shows an example of the sinewaves measured by the accelerometer and force sensors, where in red, the portion of steady state signal is analysed. The test was conducted at ten drive voltages from 0.1V up to 1V, with a 0.1V increment.

The Zhang-Zang method was applied to the acceleration and force surfaces. The acceleration surface was cut by a plane at constant acceleration, and the voltage axis found the corresponding forces common to the two surfaces. The standard Dobson method calculated and processed a set of linearised *FRFs* to extract the modal parameters. The modified-Dobson method processed the nonlinear *FRFs* generated using the steady state part of the acquired signals to extract amplitude-dependent modal parameters, as shown in Figure 20. Figure 21 and Figure 22 show the natural frequency and damping curves for the stepwise and modified-Dobson analyses. The Zhang-Zang method confirms the results of the modified-Dobson method.

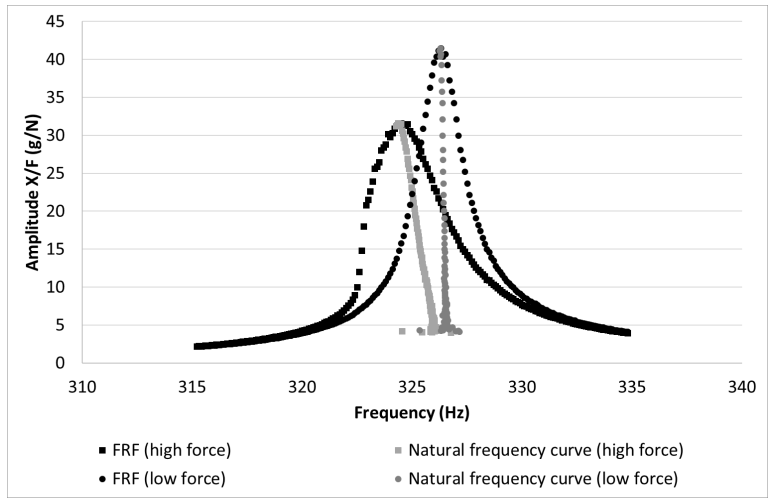

**Figure 20 Two nonlinear *FRFs* processed by the modified-Dobson method.**

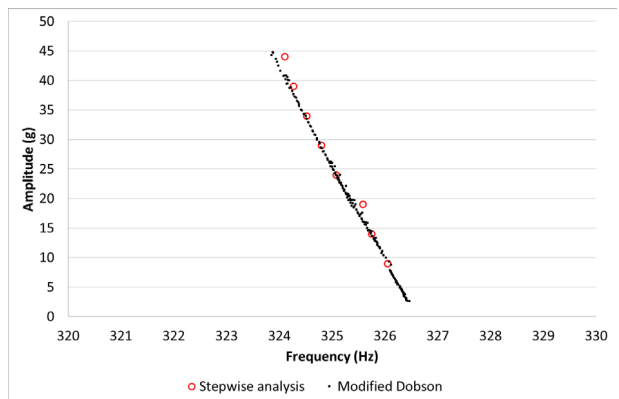

**Figure 21 Natural frequency curves compare modified-Dobson and stepwise analysis methods (note these are expressed in acceleration (g)).**

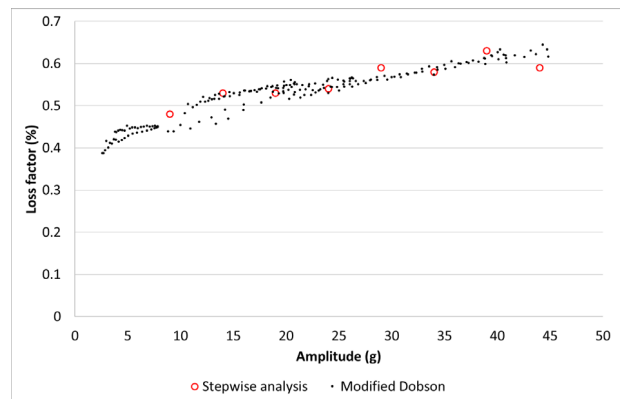

**Figure 22 Damping curves compare modified-Dobson and stepwise analysis methods.**

## 4.2   Validation of the new method for setup-2: a dumbbell setup

This first stage of the validation of the new method is based on an experiment carried out on a dumbbell setup with a shaker, as shown in Figure 22. Most details are reported in [19]. This section will resume the central aspect of that publication. The test structure was designed to have two steel masses weighing approximately 4 kg. The lap joint comprises two square aluminium sections, each connected to the mass with a single 10/32 UNF bolt and coupled to each other with two M5 bolts. Two torque levels of 6 Nm and 10 Nm were applied to the bolts to track the behaviour differences. The structure was instrumented with four single-axis accelerometers placed at the ends of the cylinders, and a force gauge was installed in the axial direction. The dumbbell was then supported by two belts and suspended in free-free conditions by elastic cords.

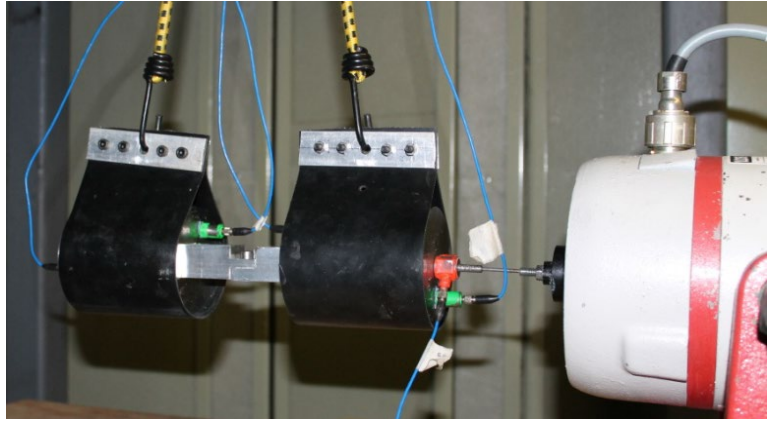

**Figure 23 Dumbbell test setup.**

The test regards the stepped sine excitation, which was carried out at pure tone excitation in a frequency range between 1200 [Hz] and 1360 [Hz] with a frequency resolution of 0.2 [Hz]. The test was carried out by controlling the excitation force, which was kept constant over the frequency range. These campaigns used several excitation forces, measured for a tightening torque of 6 Nm and 10 Nm, respectively. The measured *FRFs* showed an evident distortion of the response curves as the excitation force is increased, demonstrating the amplitude dependency of this nonlinear phenomenon under study. Figure 24 shows an example of *FRFs* measured at various excitation forces for a tightening torque 10 Nm. The *FRF* measured at 25 N is processed by the modified-Dobson method. The amplitude-dependent modal parameters were then used to generate the surface and calculate nonlinear *FRFs*. Figure 25 shows the black nonlinear *FRF* measured at 25N, with the natural frequency curve in red. Polynomials fitted the four nonlinear modal parameters as a function of the modulus of the Accelerance, $|A(\omega)| = \overline{A}$. Table 4 reports the polynomial functions used for generating the surface, recalling that those polynomials were converted to acceleration functions. The 3D surface plot is made by Accelerance (g/N), Frequency (Hz), Acceleration (g).

| Natural Frequency (Hz) | $\omega_r(\overline{A}) = 3.067 * \overline{A}^2 - 24.52 * \overline{A} + 1265.1$ |
|---|---|
| Damping loss factor (-) | $\eta_r(\overline{A}) = 0.0006 * \overline{A} + 0.0082$ |
| Modal constant - Real | $\text{RE}(_rA(\overline{A})) = -33975 * \overline{A}^2 - 136804 * \overline{A} + 891077$ |
| Modal constant - Imaginary | $\text{IM}(_rA(\overline{A})) = 42430 * \overline{A} - 26831$ |

**Table 4 Amplitude-dependent modal parameters in the form of polynomial functions.**

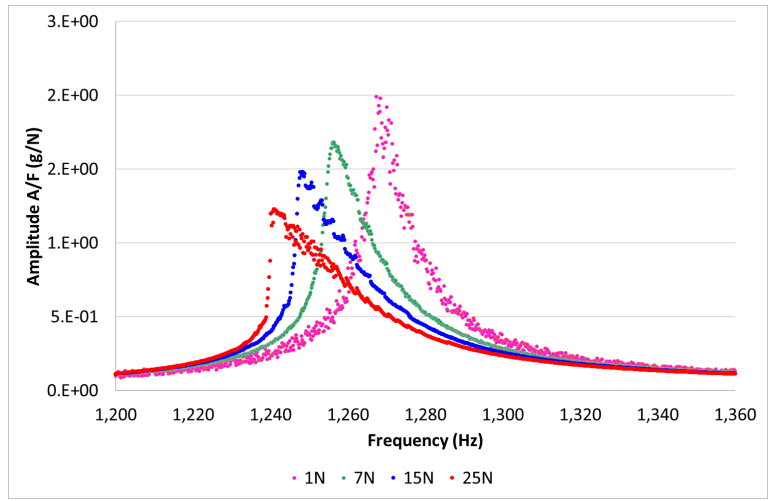

**Figure 24 Nonlinear *FRFs* at various force levels for 10Nm torque.**

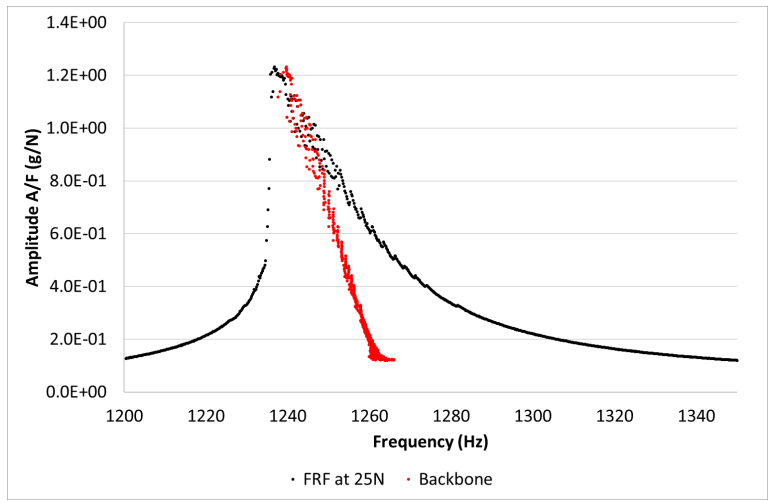

**Figure 25 Nonlinear *FRF* at 25 N in black with the natural frequency curve in red.**

The surface is graphically similar to the one shown in Figure 4 and is not reproduced here. The goal is to compare the experimental *FRF* at 25 N force with the regenerated one extracted from the surface. Furthermore, it is expected that by changing the force magnitude (and so the inclination of the force plane), one can extract the regenerated *FRFs* of 15 N and 7 N force. Figure 26 shows the targeted *FRF* at 25 N measured and regenerated *FRFs* in red and black, respectively. The overlap is rather remarkable. Moreover, the same figure shows the overlay between the *FRFs* at 15 N (blue) and 7 N (green) and the regenerated nonlinear ones. Even for these two cases, the regenerated and measured ones overlay well.

One final remark: the constant term of $\omega_r$ is not the natural frequency of the underlying linear system because the *FRF* is still mildly nonlinear. A much lower excitation force is required for extracting the linear natural frequency curve. The modified-Dobson observes the change of modal parameters from the lowest vibration points of the *FRF*, which are not always as low as desired for extracting the linear parameters.

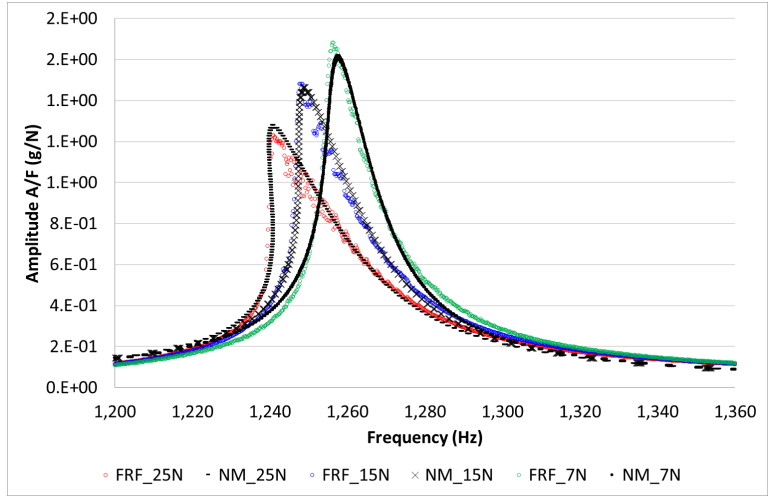

**Figure 26 Measured and regenerated nonlinear *FRFs*.**

## 4.3    Validation of the new method using setup-3: a composite blade

The third validation stage was carried out using *FRF* data measured from cantilever blades made of composite materials. The work was published in [20], and more details can be found in that manuscript. Here, a summary is given to the reader to contextualise the type of experiments executed on such test structures. The blades were designed using different layup configurations to enhance the nonlinear vibration response. Three different stacking sequences of composite materials were selected, and these were a unidirectional named (0-0), a cross-ply named (0-90) and a cross-ply named (+45, -45). The first three modes were investigated for each configuration. The measurements of the *FRFs* were carried out using a contactless excitation system called the Pulsed Air Jet system, which was developed by the author of this paper with the support of Rolls-Royce plc [21].

The excitation force was exerted onto the blade by jets of compressed air sampled by a spinning perforated disc. The rotational speed of the disc could be adjusted, and the jets' rate could be yielded to excite the resonances of the blade. The blade was mounted inside a mass block of about 40kg, where another smaller mass block was pressed utilising two large bolts, the root of the blade, as shown in Figure 27. The nozzle of the Pulsed Air Jet was directed to the corner of the blade. This exciter is contactless, meaning the excitation force exerted by the compressed air jet cannot be measured. An attempt was made to calibrate the measured pressure at the nozzle with a load cell, but the results were not accurate. Therefore, the *FRFs* presented in this paper were scaled by an arbitrary constant unit force. A single-point laser vibrometer was used to measure the response at the blade's top-left (or top-right corner), as shown by the reflective tapes in Figure 27. The *FRFs* were measured under steady state response, which meant changing the rotational speed step-by-step, with an elapsed time to settle the transient response. The acquisitions of the steady state signals were triggered by the one-pulse per revolution of an encoder mounted on the shaft of the rotor used to measure rotational speed. Such an acquisition arrangement allowed for measuring the modulus and phase of each spectral frequency point. The excitation level was changed manually by opening the compressed airflow's inlet valve to the exciter system's plenum chamber. Hence, this manual operation is another reason for scaling the force arbitrarily.

Two examples of *FRFs* are presented here. The first one was measured from a blade made of cross-ply laminate, where the second bending mode of the blade was measured. The *FRF* was processed by the modified-Dobson method, and the nonlinear modal parameters were fitted by the polynomials function of the mobility *FRF* modulus $|V(\omega)| = \overline{V}$, as shown in Table 5. Converting the mobility *FRF* in velocity is unnecessary because the scaling force was set to unity.

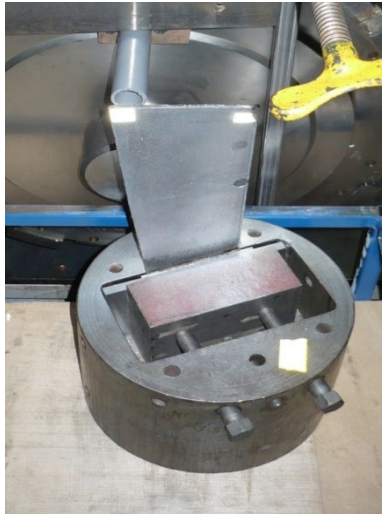

**Figure 27 Blade mounted onto a block with the nozzle of the exciter at the top-left corner.**

Figure 28 shows the measured *FRF* (grey) with the backbone extracted by the modified-Dobson method (red) and the regenerated nonlinear *FRF* (black). The comparison is remarkable, so much so that the backbone could be seen continuing up to the peak displacement of the regenerated nonlinear *FRF*.

| Natural Frequency (Hz) | $\omega_r(\bar{V}) = -2*10^8*\bar{V}^3 + 748972*\bar{V}^2 - 4057.6*\bar{V} + 286.2$ |
|---|---|
| Damping loss factor (-) | $\eta_r(\bar{V}) = -1378*\bar{V}^2 + 5.3715*\bar{V} + 0.0124$ |
| Modal constant - Real | $\mathrm{RE}(_rA(\bar{V})) = -13615*\bar{V} + 17.522$ |
| Modal constant - Imaginary | $\mathrm{IM}(_rA(\bar{V})) = -4*10^8*\bar{V}^3 - 1*10^7*\bar{V}^2 + 31053*\bar{V} + 175.6$ |

**Table 5 Amplitude-dependent modal parameters (0-90) second bending resonance**

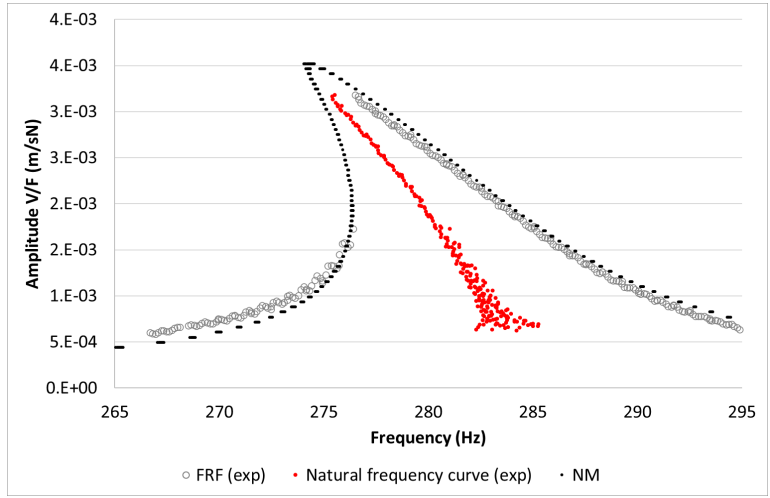

**Figure 28 Second bending mode of the composite blade (0-90).**

Given the accurate regeneration of the nonlinear *FRF*, an attempt to extend the amplitude range of the previous regeneration was yielded. Therefore, the surface *FRF* and a new force surface were simulated to extract a new nonlinear *FRF*. The *nonlinear frequency response* surface is not reproduced because finding a suitable perspective to appreciate how the surface was shaped was impossible. Figure 29 shows a new *FRF* presenting a classical Duffin oscillator shape with an additional separate loop resembling an isola [22].

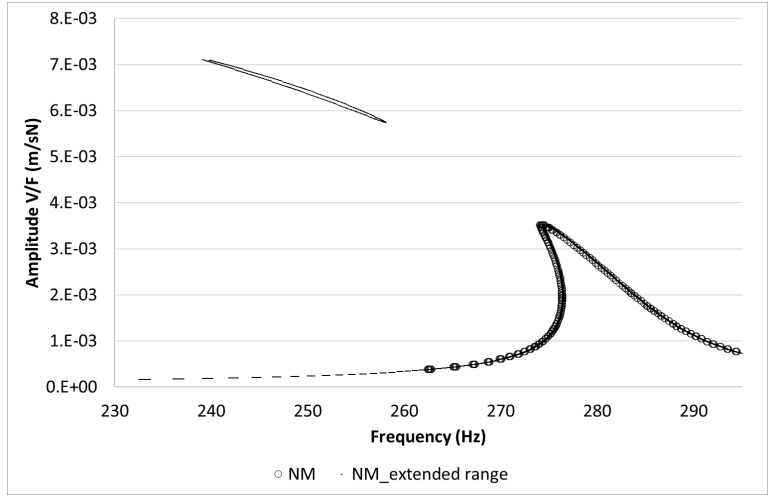

**Figure 29 Simulation with an extended amplitude range.**

The last example describes the response measured from a composite blade made of a unidirectional laminate and vibrated at its first torsional mode. The experiment, the application of the modified-Dobson method, and the generation of the surface are the same as discussed earlier. Table 6 shows the polynomial functions used for generating the surface.

| Natural Frequency (Hz) | $\omega_r(\overline{V}) = -2*10^9*\overline{V}^3 + 5*10^6*\overline{V}^2 - 1191.7*\overline{V} + 306.24$ |
|---|---|
| Damping loss factor (-) | $\eta_r(\alpha_V) = 0.016$ |
| Modal constant - Real | $\text{RE}(_rA(\overline{V})) = 6*10^9*\overline{V}^3 - 1*10^7*\overline{V}^2 + 889.38*\overline{V} - 123.46$ |
| Modal constant - Imaginary | $\text{IM}(_rA(\alpha_V)) = -8*10^9*\alpha_V^3 + 2*10^7*\alpha_V^2 - 5098.2*\alpha + 35.035$ |

**Table 6 Amplitude-dependent modal parameters (0-0) first torsional resonance**

The damping was constant over the vibration amplitude analysed, but its value had to be slightly adjusted from 0.018 to 0.016 to match the peak amplitude of the regenerated *FRF* with the experimental one. Figure 30 shows the overlay of the experimental and regenerated *FRF*s, including the experimental natural frequency curve. The overlay shows a frequency offset caused by the natural frequency curve estimation. This divergence is caused by the selection of the frequency points swept by the *sweeper*, which were from a frequency at approx. 307 Hz up to 330 Hz. The whole frequency response curve was not used because the lower branch did not produce coherent modal parameters. Unfortunately, no alternative *FRF* was inspected for such typical *S*-shaped behaviour. Nonetheless, the overlay shows very similar shapes.

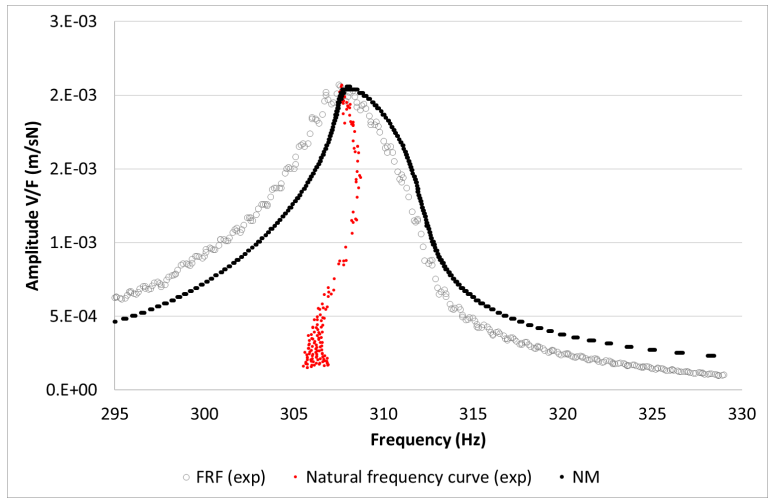

**Figure 30 Measured (grey) and regenerated (black) nonlinear *FRFs*.**

# 5 Discussions

In the introduction, the manuscript postulated the following hypothesis:

*A nonlinear frequency response surface, generated by a waterfall of linear FRFs, is the solution space of nonlinear FRFs, which can be evaluated by any force plane cutting across the response surface.*

The paper demonstrated how that surface could be built. The first option uses a static nonlinear receptance, and the second uses amplitude-dependent modal parameters. Even though the proposed analysis looks like the one proposed by Özgüven [12], the main difference lies in using linear *FRFs* to generate a *nonlinear frequency response* surface. Hence, the entire analysis is computationally cheap, avoiding sophisticated numerical techniques. The nonlinear drive frequencies, $\omega_{NL}$, of the nonlinear *FRF* are unique solutions of the *nonlinear frequency response* surface for an arbitrary force place. The final remark is about using the definition of linear operator, the Frequency Response Function, to label the nonlinear frequency responses. The authors believe it is possible to use the exact definition because the nonlinear response amplitudes are evaluated at unique drive frequencies, meaning there are no multiple solutions for one drive frequency.

Regarding the modal identification, the modified-Dobson formulation allowed the calculation of the nonlinear modal parameters. It must be said that this method is not exclusive, but it overcomes the limitation of accessing a few *FRFs* rather than many. The experimental analysis proved that a well-isolated resonance could be identified, and nonlinear *FRFs* could be regenerated and compared to the experimental ones. The challenge is when the modes become closer, whereby the current formulation fails to return reliable nonlinear modal parameters. It was already stated in section 3.3 that the close modes analysis was addressed for the linear modal analysis, and a similar approach can be extended to the nonlinear one. This new method is proven for single-input, single-output (SISO) testing. Upcoming papers will present more detailed analyses of the applicability of the modified-Dobson method, even focussing on the extension to the single-input, multi-output (SIMO) tests.

The new analysis method aims to unlock some constraints observed in (i) the experimental modal analysis and (ii) the finite element analysis.

The first constraint was about nonlinear modal parameters identified, which could not be readily used for regenerating the measured *FRF*. Over the past years, many techniques have been developed to simulate nonlinear *FRFs* using experimental parameters. However, no matter how excellent those techniques are, they cannot be easily standardised for industrial use. The new analysis method is based on a linear theory already embedded in industry standards. Therefore, a few extra implementations in existing modal analysis toolsets are required to yield the regeneration of nonlinear *FRFs*. Furthermore, the new analysis method establishes modal parameter validation, which is possible for linear modal analysis, and is required to evaluate if the regenerated nonlinear *FRF* matched, or not, the measured one. An example of validation of modal parameters was presented in section 4.3, where the second test case required a minor adjustment of the damping identified to regenerate and match a nonlinear *FRF* with the experimental one.

The second constraint concerns the computational cost of running nonlinear time-domain simulations using analytical and finite element methods. This manuscript showed that a stress model can generate nonlinear *FRFs* by evaluating a force-displacement curve sufficient to build a *nonlinear frequency response* surface. A nonlinear damping formulation would also be required by carrying hysteresis simulations, as done in [7]. Once the nonlinear stiffness and damping are evaluated, an engineer can simulate forced nonlinear vibrations under steady state conditions for the mode under study. An example based on the finite element method was presented in Appendix A2.

# 6 Conclusions

The primary objective of this manuscript was to enable engineers to analyse, regenerate and compare nonlinear *FRFs*, following a similar process established for linear modal analysis. This goal was achieved for *FRFs* measured under steady-state vibrations, smooth nonlinearities and well-separated modes. These conditions are not always possible for mechanical systems. Nonetheless, it was essential to (i) establish a new analytical framework for processing nonlinear frequency response, (ii) regenerate, and (iii) compare them. The manuscript presented a modified-Dobson method for extracting nonlinear modal parameters. However, any existing experimental technique is valid if the natural frequencies, damping and modal constant nonlinear curves can be evaluated.

## Acknowledgements

I wish to acknowledge Prof. David Ewins, with whom I exchanged uncountable thoughts about modal analysis. I would also like to acknowledge Mr. Jip van Tiggelen for the support offered during the write-up of this manuscript.

# APEENDIX

## A1    Block diagram to operate the new analysis method

The block diagram presented in Figure 31 shows a stepwise approach of the proposed new analysis method. Note that the look-up table requires the parameters' curves to create much finer steps than an analysis can typically produce. Hence, many interpolation points can be created between the ones identified by the modified-Dobson method. Remembering that the look-up table is also limited to the maximum vibration amplitude analysed is also good, for this case, the range cannot extended as it is possible for polynomial functions.

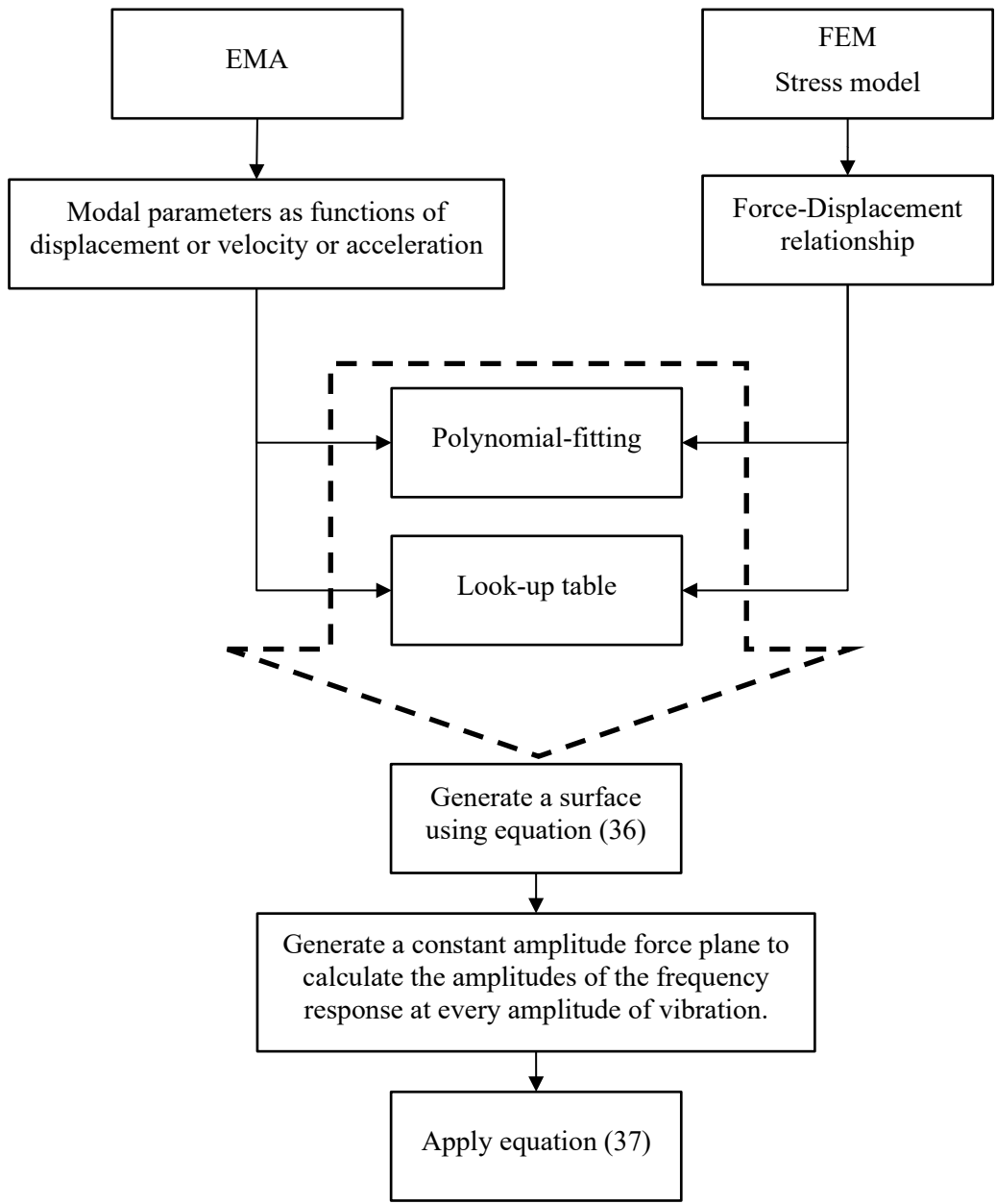

**Figure 31 Block diagram for the application of the new analysis method.**

## A2 Application to FE models

This final sub-section will evaluate if the new theoretical formulation can be extended from the two-DOF system to finite element simulations. The motivation follows the research work by Singh et al. [7], who have been publishing on a novel technique called Quasi-Static Modal Analysis. The research is very interesting because it can extract modal parameters, such as natural frequency and damping, which are amplitude-dependent. The simulations are kept within the finite element solution space without intruding, for instance, by developing external plug-ins. The novelty of their research is to calculate the change of internal stress under a static given modal shape and then to calculate the equivalent natural frequency at that given state of deformation. The equilibrium point around which the modal analysis is carried out changes for any given deflected state imposed on the model. The formulation in this paper is equivalent to theirs, but the modal space is completely bypassed using the direct derivation of the response model rather than the modal model. Both methods are equivalent and serve different purposes. The current new method is mainly intended to calculate the force vibrations under steady-state conditions. Therefore, the following two examples will show how forced vibrations can be directly calculated for the first vibration mode for two simple beam structures. It is essential to understand that the two examples are purely hypothetical; therefore, the models do not represent physical structures except geometry.

### A2.1 Cantilever beam model

A simple prismatic cantilever beam made of mild steel, the property of which was taken from tabulated spreadsheets, is modelled by finite element, as shown in Figure 32. An applied force deflects the beam at the tip of the beam. A force-displacement nonlinear curve is yielded by the applied force, as shown in Figure 33. A polynomial function of the third order fitted the force-displacement curve. The polynomial function was then converted to a static transfer function, equation (40), which is required to calculate the nonlinear *FRF* surface. After carrying out the static analysis, modal analysis was yielded to calculate the first natural mode of the cantilever beam, the frequency of which was 41.59 Hz. That frequency and the equivalent linear stiffness, from equation (40), was used for calculating the equivalent mass. The viscous damping was assumed to be linear with a value $c$ = 5 (Ns/m). A single degree of freedom response model was used to generate the nonlinear *FRF* surface, and two force surfaces were also generated to extract a linear and nonlinear transfer function.

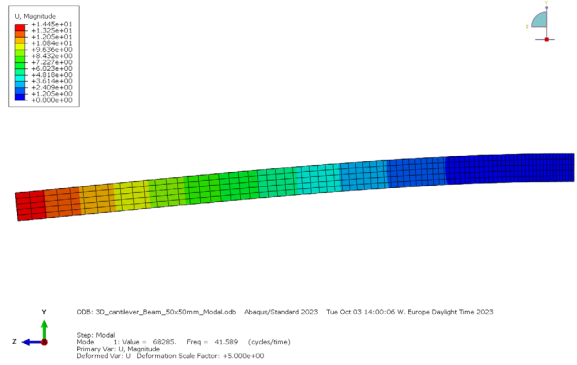
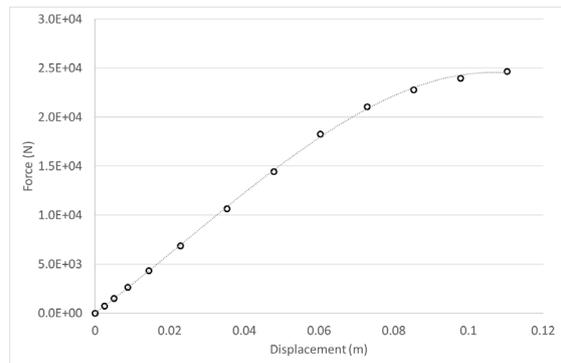

**Figure 32 FE model of a cantilever beam.**       **Figure 33 Force-Displacement curve.**

Figure 34 shows both the linear and nonlinear *FRFs* simulated. The nonlinear *FRF* shows a hardening and softening behaviour, given the stress-strain relationship evaluated by the static deflection.

$$\alpha(X) = \frac{X}{F} = \frac{1}{-2*10^7 * X^2 + 1*10^6 * X + 287827} \tag{40}$$

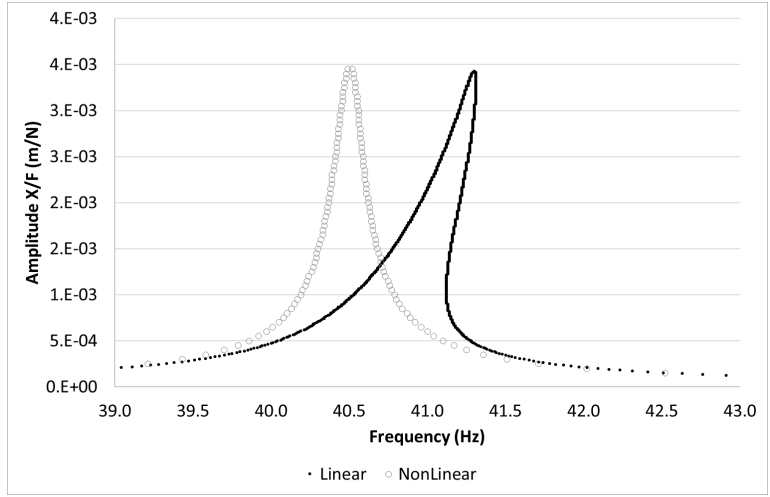

**Figure 34 Linear and nonlinear FRF obtained from the beam model.**

## A2.2   Two-jointed beams model

The second test case is a two-jointed beams FE model subjected to a fully reverse load in a three-point bending test simulation. The beams are mild steel with material properties selected from tabulated spreadsheets. A bolt joins the beams, and the joined interfaces are preloaded to mimic the torque load imposed to the bolt. Standard ABAQUS nonlinear contact elements are used for the interface and Coulomb friction coefficient, $\mu = 0.4$, specified at those elements. A three-point bending test simulation is generated by fixing the extreme points of the beam and applying a force in the middle. The beam is subjected to fully reverse cycle load to simulate a hysteresis curve. Figure 35 shows the test structure as modelled in FEM, where the two extreme points are fixed (rotation enabled), and a force is applied to the middle. Figure 36 shows the initial part of a hysteresis curve calculated and a slight nonlinear deflection beyond the 0.3 mm displacement.

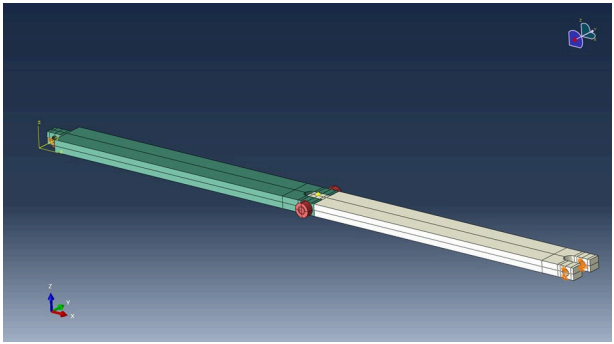

**Figure 35 Two-jointed beam model with fixed boundary conditions and force application in the middle.**

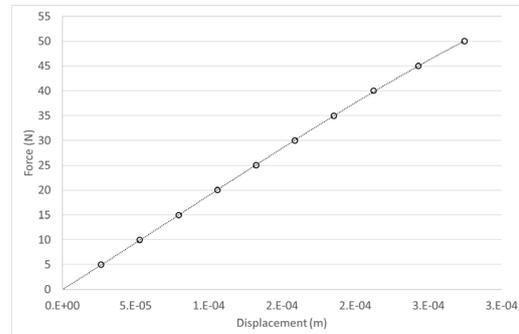

**Figure 36 A portion of the hysteresis curve calculated from the fully reverse loading cycle.**

The force-displacement relationship, shown in Figure 36, is converted to a static receptance expressed by equation (41). The linear modal analysis is carried out to calculate the first resonant frequency, 195 Hz. A linear viscous damping is used with $c = 1$ (Ns/m).

$$\alpha(X) = \frac{X}{F} = \frac{1}{-4*10^{11}*X^2 + 1*10^8*X + 183345} \qquad (41)$$

Figure 37 shows three FRFs extracted by the simulated surface at the linear and nonlinear amplitudes. The *FRF* with low force level shows a mild hardening behaviour, which becomes softening behaviour at an higher force amplitude. However, joints exhibit both stiffness and damping nonlinearity. For this example, hysteresis was used for the nonlinear stiffness behaviour without evaluating the damping. A pseudo nonlinear quadratic damping function is assumed, as typically taken in a textbook of nonlinear dynamics [4]. Therefore, the surface *FRF* is regenerated using equation (41) and damping as a function of the displacement described by equation(42).

$$d(X) = d_{LIN} X + d_{NL}^2 |X|^2 \tag{42}$$

A new FRF surface is generated, including stiffness and damping nonlinearity. Five force surfaces are used to extract the nonlinear *FRFs*, as shown in Figure 38. This virtual experiment was yielded to observe how the combination of both nonlinearities would appear in a transfer function since the only stiffness nonlinearity might generate a difficult-to-measure *FRF*. Adding damping nonlinearity generated *FRFs* that are more similar to the ones typically measured in jointed structures.

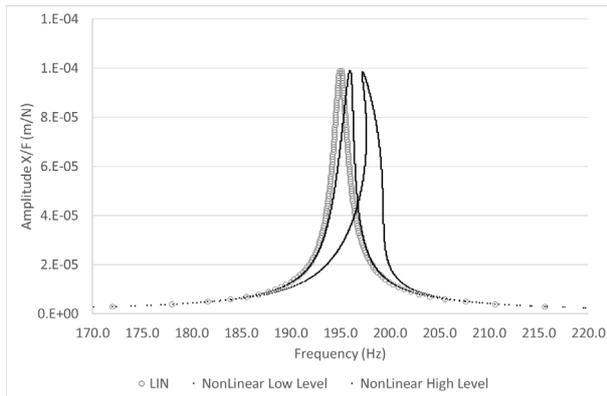

**Figure 37 Three *FRFs*, one linear, one low level of force and one high level of force.**

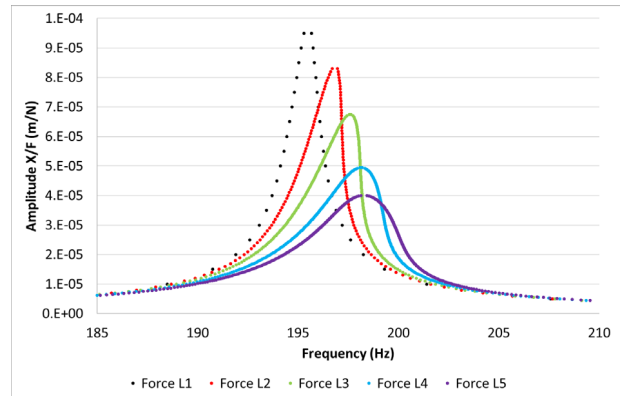

**Figure 38 Nonlinear *FRFs* at various force levels simulated with both stiffness and damping nonlinearity.**

## A3    Validation of the modified-Dobson method

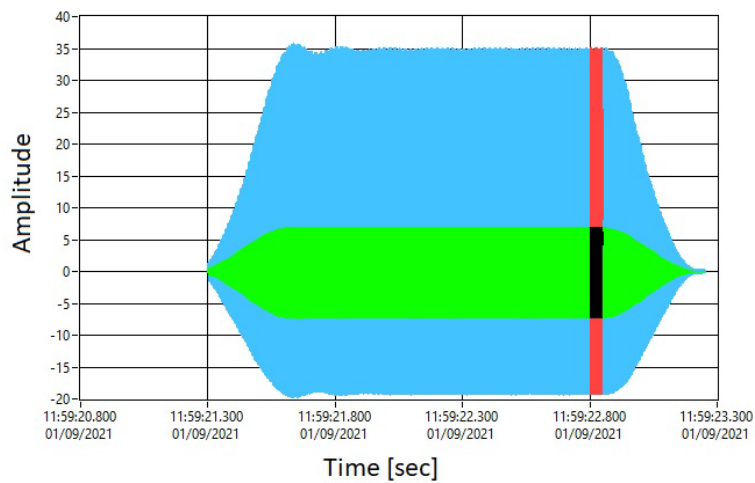

**Figure 39 An example of a signal measured by the accelerometer and force sensors. In red, the steady state part is analysed.**